\documentclass[twocolumn]{openjournal}

\usepackage{orcidlink}
\usepackage[T1]{fontenc}
\usepackage{ae,aecompl}
\usepackage{graphicx}	
\usepackage{amsmath}	
\usepackage{amssymb}	

\usepackage{natbib}
 
\usepackage{hyperref}
\hypersetup{
	colorlinks=true,
	urlcolor=blue,    
	linkcolor=black,  
	citecolor=blue,   
}

\defcitealias{Pittordis_2018}{PS18}
\defcitealias{Pittordis_2019}{PS19}
\defcitealias{Badry_2021}{ERH}
\defcitealias{Banik_2018_Centauri}{BZ18}

\hypersetup{citecolor=blue, 
            linkcolor=red, 
            menucolor=blue, 
            urlcolor=blue}  

\newcommand{\cm}{{~\rm cm}}

\newcommand{\km}{{~\rm km}}
\newcommand{\s}{{~\rm s}}

\newcommand{\g}{{~\rm g}}

\newcommand{\K}{{~\rm K}}

\newcommand{\yr}{{~\rm yr}}
\newcommand{\yrs}{{~\rm yrs}}

\newcommand{\pc}{{~\rm pc}}

\newcommand{\AU}{{~\rm AU}}

\newcommand{\days}{{~\rm days}}

\begin{document}

\title{Jet-Driven Formation of Bipolar Rings in Planetary Nebulae: Numerical Simulations Inspired by NGC 1514}

\author{Muhammad Akashi\,\orcidlink{0000-0001-7233-6871}}
\affiliation{Kinneret College on the Sea of Galilee, Samakh 15132, Israel}
\affiliation{Department of Physics, Technion - Israel Institute of Technology, Haifa, 3200003, Israel; akashi@technion.ac.il}

\author{Ealeal Bear} 
\affiliation{Department of Physics, Technion - Israel Institute of Technology, Haifa, 3200003, Israel; ealeal44@technion.ac.il}

\author{Noam Soker\,\orcidlink{0000-0003-0375-8987}} 
\affiliation{Department of Physics, Technion - Israel Institute of Technology, Haifa, 3200003, Israel; soker@physics.technion.ac.il}

\begin{abstract}
We conduct three-dimensional hydrodynamical simulations of jets launched into a dense shell, reproducing two rings in a bipolar structure that resemble the two dusty rings of the planetary nebula (PN) NGC 1514. The scenario we simulate assumes that a strong binary interaction enhanced the mass loss rate from the asymptotic giant branch (AGB) stellar progenitor of NGC 1514, and shortly thereafter, the main-sequence companion accreted mass from the AGB star, launching a pair of jets. We find that flows with negligible radiative losses produce prominent rings, as observed in the infrared in NGC 1514. In contrast, when radiative cooling is significant, the rings are thin and faint. Our results reinforce the prevailing notion that jets play a substantial role in shaping planetary nebulae (PNe). More generally, as the binary companion to the central star of NGC 1514 avoided common envelope evolution, our results suggest that jets play a major role in many binary systems experiencing stable mass transfer at high rates. This conclusion complements the view that jets play a significant role in unstable mass transfer, specifically in common envelope evolution. Studies of strongly interacting binary systems, whether stable or not, should include jets. If jets continue to be active after ring formation, the outcomes are circum-jet rings, as observed in some other PNe and core-collapse supernova remnants.
\end{abstract}  


\section{Introduction}
\label{sec:Introduction}

Pairs of structural features on opposite sides of the center are very common morphological features of planetary nebulae (PNe). The two opposite structural features in a pair might be 
faint zones fully closed by bright rims, termed bubbles, or partially closed by a rim and termed lobes, dense (bright) clumps (`ansae'), a narrow opening in the PN shell that are termed nozzles, bright arcs (rims), rings, and protrusions with a base smaller than the PN main shell and a cross-section that decreases outward, termed ears. The line connecting two opposite structural features is a symmetry axis. Multipolar PNe are those with two or more symmetry axes, and as the many papers on this topic show, attract lots of attention (e.g., \citealt{Manchadoetal1996b, SahaiTrauger1998, Sahai2000, Harmanetal2004, Kwoketal2010, Sahaietal20112011, Chongetal2012, Sabinetal2012, Velazquez2012, Clarketal2013, Guerreroetal2013, Guerreroetal2020, Steffenetal2013, Hsiaetal2014, Rubioetal2015, AkrasGonclves2016, Huangetal2016, GomezGonzalezetal2020, RechyGarciaetal2020, Bandyopadhyayetal2023, GomezMunozetal2023, MoragaBaezetal2023, Wenetal2023, Wenetal2024, Goldetal2024, Kwok2024, AvitanSoker2025, Soker2025Bright}), with some multipolar pre-PNe (e.g., \citealt{Hrivnaketal1999, Sahaietal2005Starfish, Trammelletal2002}). The PN NGC~1514 that we study here is a multipolar PN, but rather than having two or more similar pairs of lobes, it has a messy morphology.   

We adopt the view that pairs of jets that the binary system progenitor of the PN launches shape symmetry axes (e.g.,  \citealt{Morris1987, Soker1990AJ, SahaiTrauger1998, AkashiSoker2018,   EstrellaTrujilloetal2019, Tafoyaetal2019, Balicketal2020, RechyGarciaetal2020, GarciaSeguraetal2021, GarciaSeguraetal2022, Clairmontetal2022, Danehkar2022, MoragaBaezetal2023, Ablimit2024, Derlopaetal2024, Mirandaetal2024, Sahaietal2024} for a partial list; \citealt{Baanetal2021} discuss an alternative scenario). 
Other processes play a role in shaping the PN and influencing the launch of jets. One is a common envelope evolution, mainly of a main sequence star orbiting inside the envelope of an asymptotic giant branch (AGB) star; in rare cases the primary star is a red giant branch (RGB) star (e.g., \citealt{Hillwigetal2017, Sahaietal2017, Jonesetal2020, Jonesetal2022, Jonesetal2023}), and a secondary white dwarf is also possible. Another group of processes is those that change the direction of the jets during the PN formation phase, leading to precession of the jets' axis, as observed in some PNe (e.g., \citealt{Guerreroetal1998, Mirandaetal1998, Sahaietal2005, Boffinetal2012, Sowickaetal2017, RechyGarciaetal2019, Guerreoetal2021, Clairmontetal2022}), or to abrupt change in direction as required to explain multipolar PNe.  

In some cases, the two opposite sides of a symmetry axis are highly asymmetrical in size, shape, distance from the center, or not being at $180^\circ$ to each other. In extreme cases, there is no point symmetry, i.e., a messy PN. Some structural features in messy PNe are bubbles, lobes, or ears that suggest jets shaped them. Such is the messy inner structure of PN NGC~1514 (PN~G165.5-15.2), which has a large departure from axisymmetry (e.g., \citealt{Balick1987}), including in its kinematics (e.g., \citealt{MuthuAnandarao2003}). \cite{Soker2004triple} and \cite{BearSoker2017Triple} attribute PN messy morphologies to triple star interaction. While NGC~1514 is a messy PN in the visible band that reveals the inner structure, infrared (IR) observations revealed two outer rings that are more symmetric (e.g., \citealt{Ressleretal2010}). \cite{Ressleretal2025} conducted new JWST mid-infrared imaging and spectroscopy of NGC~1514 and presented a thorough study of the outer two rings; that study motivated our study.

In Section \ref{sec:NGC1514} we describe the structure of NGC~1514 and the morphological features we attribute to jets. 
We claim at least three jet-launching shaping episodes. In Section \ref{sec:outrings}, we conduct simulations that show the formation of the IR outer rings from jets. 
We summarize in Section \ref{sec:Summary} and discuss the wider implications of our study,

\section{The multiple-jet-shaped structures}
\label{sec:NGC1514}

\subsection{The morphology of NGC 1514}
\label{subsec:Morphology}

The recent thorough analysis of JWST observations of the PN NGC 1514 by \cite{Ressleretal2025} motivates this study. We aim to demonstrate that jets can shape the outer rings, thereby relating them to the different jets that shaped the inner region. 

The classification of NGC 1514 is complex. 
\cite{Chuetal1987} described NGC 1514 as a multiple-shell PN. They classify it as type IIp, where "II" denotes a multiple-shell planetary nebula with attached shells and  "p" denotes a peculiar and irregular outer shell, sometimes filamentary. 
\cite{Ressleretal2010} reported the discovery of a pair of axisymmetric rings observed in the mid-IR (WISE all-sky mid-infrared survey); the rings' emission is dominated by dust emission.
\cite{Ressleretal2010} estimated the rings to have a diameter of $\approx 0.2\pc$ separated by $0.05\pc$ with a temperature of $160 \K$.
\cite{Ressleretal2025}  defines it as a “multiple shells, point symmetry and internal structure” due to what they classify as a "lumpy bright inner shell"  which contains complex structures and a "diffuse faint outer shell". 
The morphology features of such an elliptical outer shape with multiple inner bubbles suggest a binary interaction (e.g., \citealt{MuthuAnandarao2003, Alleretal2015,Jonesetal2017}.)

\cite{Kohoutek1967} suggested for the first time that the central object in PN 1514, named  $BD+30^{\circ}623$, is composed of two components. Many studies have been done since then and \cite{Alleretal2015} later suggested the following binary parameters for NGC 1514: a horizontal branch A0 star of $M_{\rm cool} = 0.55 \pm 0.02 M_{\odot}$, or an alternative post main sequence star of mass $M_{\rm cool} = 2.9 \pm 0.5 M_{\odot}$ (based on the evolutionary tracks of \citealt{Girardietal2000}), and sdO star of $M_{\rm hot}= 0.56 \pm 0.03 M_{\odot}$. \cite{Jonesetal2017} based on the observed radial velocity amplitudes ruled out the low mass of the cool component (horizontal branch star) and claimed that the mass ratio is $q \gtrsim 2$ and suggested an alternative post main sequence star of mass $M_{\rm cool} \simeq 3M_{\odot}$. They found that the system is highly eccentric $e \simeq 0.5$, and has a long period of $P \simeq 3300 \days$. An estimate based on the inclination for the binary along with the measured eccentricity and radial velocity amplitudes result in $M_{\rm cool} = 2.3 \pm 0.8 M_{\odot}$ and $M_{\rm hot} = 0.9 \pm 0.7 M_{\odot}$ (for more details see \citealt{Jonesetal2017} and references within).

\begin{figure*}
\begin{center}
\includegraphics[trim=0cm 9.05cm 0cm 0cm, clip, width=1.0\textwidth]
{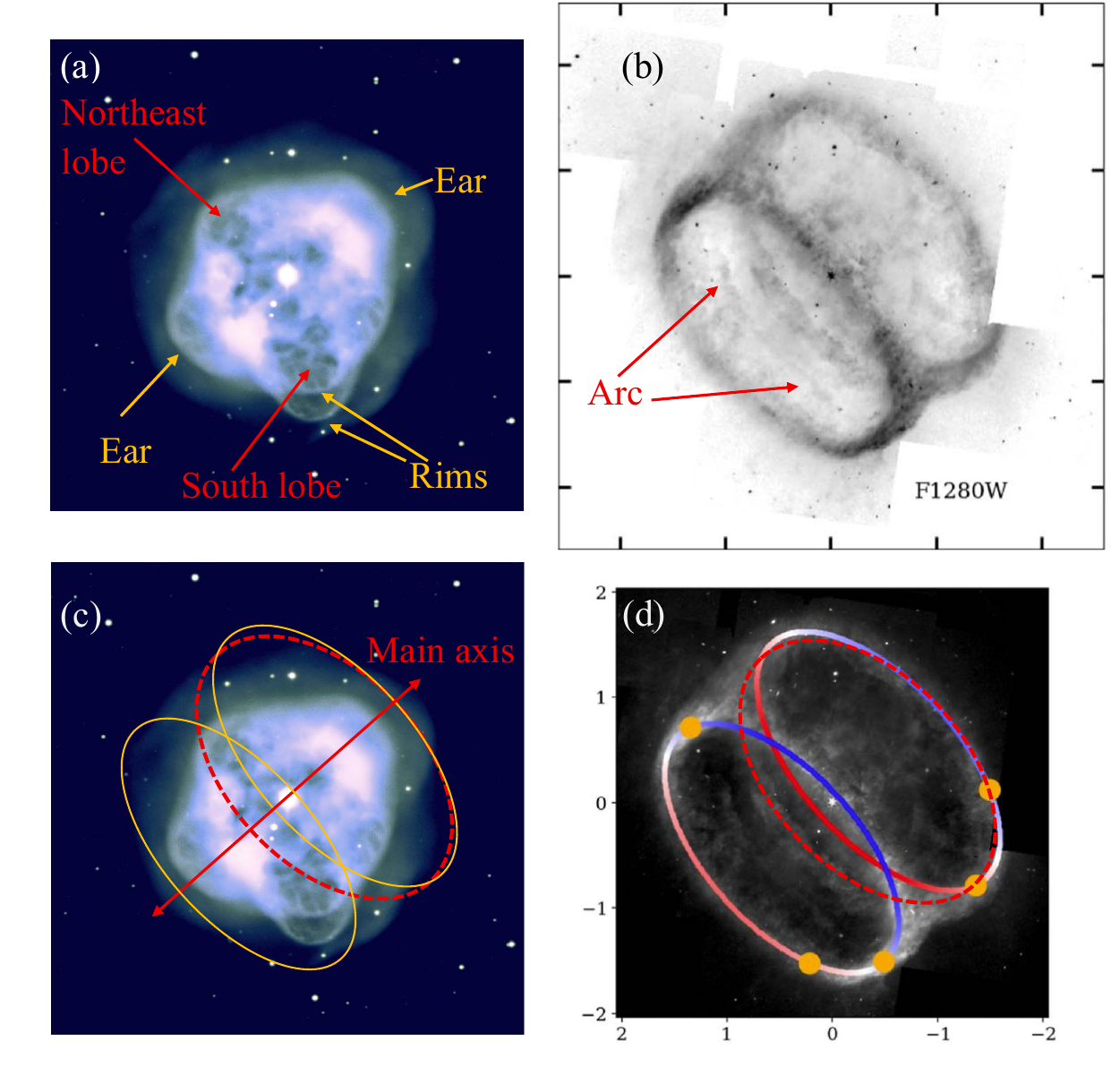}
\caption{Images of NGC 1514. All images are on the same scale with ticks an arcmin apart; north is up and east is to the left. (a) An image adapted from \cite{Jonesetal2017} who construct the image from archival O\textsc{[iii]} and
H$\alpha$+N\textsc{[ii]} images taken with the Isaac Newton Telescope Wide Field
Camera. We added all the marks. (b) A JWST $12.8 \mu$m image adapted from \cite{Ressleretal2025}. The origin $(0,0)$ (central ticks) is at the central star. We point at an arc. (c) The visible-bands image from panel a, with the ellipses from panel d, and a double-ended arrow through the center of the ellipses and the central star; this is the main axis. (d) An image adapted from \cite{Ressleretal2025} where they mark two almost identical ellipses (in solid red and blue, signify Doppler shifts). We added an alternative ellipse in the northwest, marked with a dashed red line. Our suggested ellipse signifies the departure from mirror symmetry of the two main bipolar rings.  }
\label{fig:NGC1514iamges}
\end{center}
\end{figure*}

In panels a and c of Figure~\ref{fig:NGC1514iamges}, we present an image in visible bands that we adapted from \cite{Jonesetal2017}. The low-surface brightness protrusion to the south contains two bright rims. In some other astrophysical objects, multiple jets can form multiple rims in a lobe. One example is the PN~KjPn~8 that has three rims on one of its elongated lobes, and these are attributed to shaping by jets (e.g., \citealt{Lopezetal2000}). A clearer example is the AGN jets of Hercules A, where radio emission directly reveals the jets and their rims (e.g., \citealt{Ubertosietal2025}). We suggest two jet-launching episodes that formed the two southern rims of NGC~1514.   

The opposite lobe to the southern lobe is in the northeast, rather than in the north. More generally, the inner part (visible image) of NGC 1514 lacks any clear symmetry, nor mirror symmetry, nor axial symmetry. We attributed this `messy morphology' to triple-star interaction \citep{BearSoker2017Triple}. 

\cite{Ressleretal2025} focused on the distinctive pair of rings, which is only visible in the IR and dominated by emission from small grain dust. In Section~\ref{sec:outrings} we show that jets can form the rings' morphology; these jets are earlier than those which shaped the inner structure of NGC 1514. Here, we first address the question of whether we can see signatures of departure from axisymmetry in the outer two rings that \cite{Ressleretal2025} analyzed. In panels b and d of Figure \ref{fig:NGC1514iamges}, we present two images from \cite{Ressleretal2025}, a JWST IR image in $12.8 \mu$m in panel b, and the rings with two ellipses (solid lines) that \cite{Ressleretal2025} drew, and one ellipse that we drew (dashed-red line). 
In panel b, we identify an arc. This shows that the outer dusty structure contains more than two smooth and symmetrical rings. 

\begin{figure}
\begin{center}
\includegraphics[trim=0cm 8.5cm 0cm 0cm, clip, width=0.45\textwidth]
{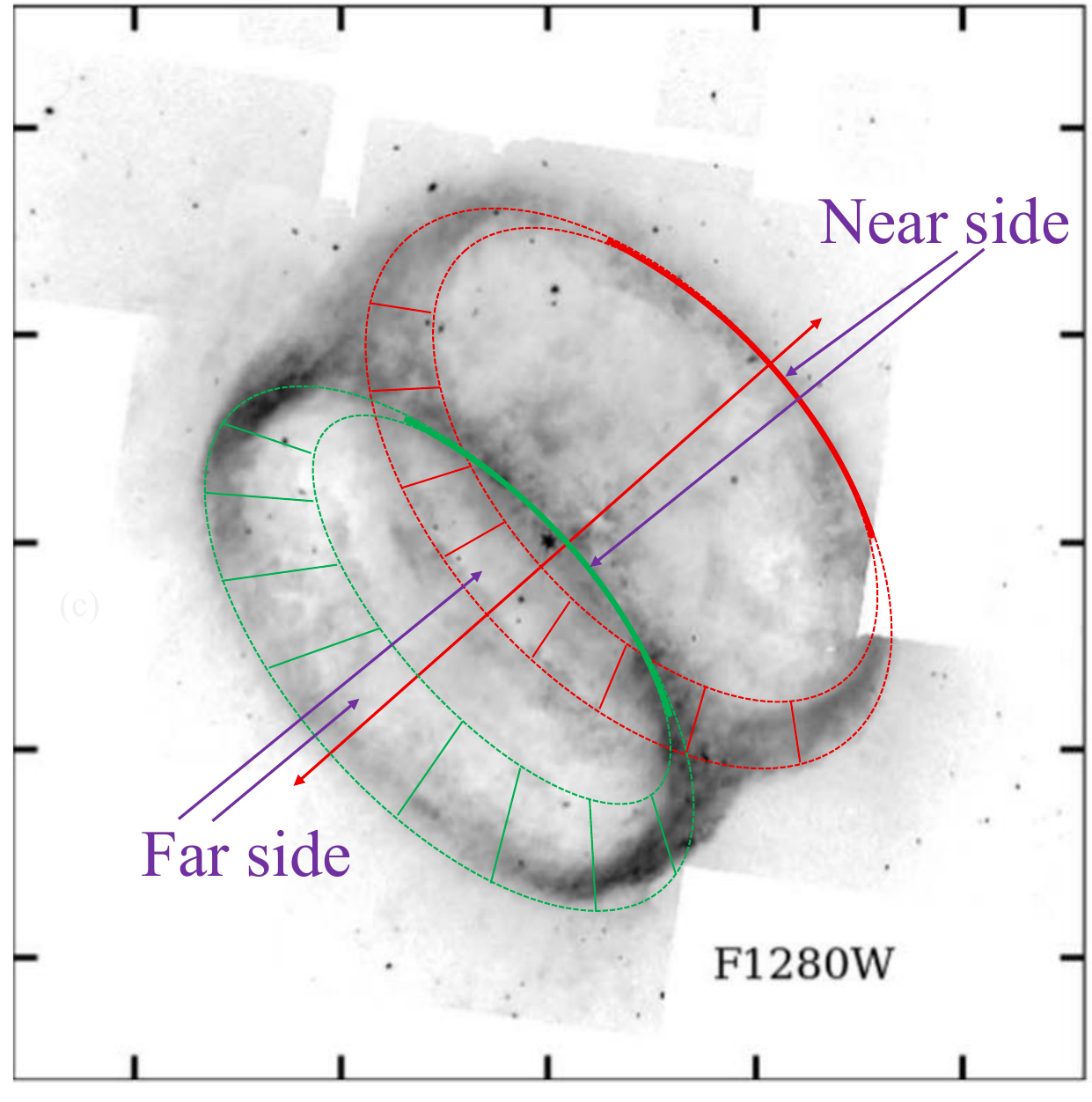}
\caption{A JWST $12.8 \mu$m image adapted from \cite{Ressleretal2025} with our suggestion of a possibility of two bipolar rings that are much wider on the far side than on the near side. }
\label{fig:WideRings}
\end{center}
\end{figure}

While \cite{Ressleretal2025} identify two rings of approximately the same size and shape, we prefer one of two different possibilities that make the rings' structure less symmetric, which is more compatible with the messy morphology of the central structure of NGC 1514. (a) The main ring in the northwest is the ring that we draw by the dashed red line. In this case, the northwest ring is much wider than the southwest ring. (b) The two rings are not axisymmetric, but rather wider on the far side (redshifted side). They indeed are similar to each other, but lack pure axisymmetry. The wider part of the northwest ring is bounded by the ring that \cite{Ressleretal2025} mark and by the ring we mark. The southeast ring is bound by the ring that \cite{Ressleretal2025}  mark and the arc that we identify. We present this possible structure in Figure~\ref{fig:WideRings}. Overall, we raise here the possibility that the outer bipolar pair of rings lacks either axial symmetry or mirror symmetry, or both.   

\subsection{A plausible scenario}
\label{subsec:Scenario}

According to our interpretation, the outer bipolar ring structure lacks pure mirror symmetry (panels c and d of Figure~\ref{fig:NGC1514iamges}), or pure axial symmetry (Figure~\ref{fig:WideRings}), or even lacks both. The lack of mirror symmetry may result from the presence of a tertiary body, stellar or sub-stellar, orbiting the binary system or one of the stars, but in a plane inclined to the binary orbital plane. The lack of axial symmetry might result from an eccentric orbit, as this binary has. Suppose most mass is lost at periastron passages of the companion around the AGB progenitor of the PN. In that case, the eccentric orbit means the closest binary approach occurs in a specific direction from the AGB star, implying that the system can lose more mass on one side of the equatorial plane. This might explain the widening of the rings towards that direction, as we draw in Figure~\ref{fig:WideRings}, if our interpretation of Figure~\ref{fig:WideRings} holds. 

Considering the `messy morphology' of the inner region that we attribute to triple star interaction \citep{BearSoker2017Triple} and the departure from symmetry of the outer bipolar ring structure, we propose the following plausible scenario of three phases to explain the shaping of NGC 1514. 

In the first phase of the strong binary interaction, the main effect of the stellar companion is to enhance mass loss from the AGB progenitor of NGC 1514. This can be caused by spinning up the AGB star as the orbit shrinks and as the periastron passage separation decreases. During that phase, the mass loss might lack axial symmetry in the case of an eccentric orbit. Namely, the mass loss rate per unit solid angle and possibly the outflow velocity vary with direction. 

In the second phase, the secondary star accretes mass and launches two opposite jets. The interaction is not purely axially symmetric due to several possible processes. ($i$) The dense shell that the AGB star ejected in the first phase might lack axial symmetry. ($ii$) The jet-launching companion is on an eccentric orbit. ($iii$) The jet-launching period is shorter, or not much longer, than the orbital period, as we suggest in Section \ref{subsubsec:RadiativeCooling}. ($iv$) A tertiary object with an inclined orbit might influence the launching of the jets, particularly making them unequal. 
In Section \ref{sec:outrings}, we present simulations of the interaction of the jets with the dense shell that the binary ejected in the first phase of the strong interaction. Our simulations in this initial study disregard the non-axially symmetric processes and the orbit of the jet-launching companion around the center of mass, as we aim only to demonstrate that jets can, in principle, form bipolar rings. 

The third phase of the interaction formed the messy inner structure of NGC 1514. In the third phase of the interaction, in addition to the accretion of AGB mass by the companion, it also accreted a third sub-stellar object, i.e., a massive planet or a brown dwarf. The pre-merger orbital plane of the tertiary around the secondary is highly inclined to the orbital plane of the binary stellar system. This makes the system more prone to dynamical instabilities. The accreted tertiary mass has its angular momentum highly inclined to the stellar binary orbital angular momentum, leading to the launching of inclined jets. In a previous study \citep{AkashiSoker2017inc}, we simulated such inclined jets and demonstrated that they indeed form messy PNe, with two opposite lobes that are not exactly at $180^\circ$ and are unequal to each other.
The two rims in the southern lobe of the inner structure suggest that there were two jet-launching episodes as a result of the merger of the sub-stellar companion into the secondary envelope.  

The densities of massive planets, a few times or more the mass of Jupiter, and brown dwarfs are larger than the density of a main-sequence star, in particular after the main-sequence star has accreted some mass from the AGB star and expanded. Therefore, it is unlikely for the main-sequence companion of the AGB star to destroy the sub-stellar object tidally. We assume, instead, that during the merger process, the sub-stellar object ejects a mass $M_d = \eta_d M_{\rm p}$ into an accretion disk around the main-sequence star, where $M_{\rm p}$ is the mass of the sub-stellar object and $\eta_d \ll 1$. Such a mass transfer might be the outflow of bound material through the second Lagrange point (behind the sub-stellar object) as it spirals into the main sequence envelope. The accretion disk launches a fraction $\eta_j$ of this mass into the jets, and at about the escape velocity from the main sequence star $v_j \simeq 600 \km \s^{-1}$. We consider the jets to interact with a nebular mass $M_I$ (which is a fraction of the total nebula as the jets are collimated), expanding with a velocity of $v_I \simeq 30 \km \s^{-1}$. The ratio of the kinetic energy of the jets to that of the nebular mass they interact with is 
\begin{equation}
\begin{split}
     f_{\rm j,I} = 2 &
     \left( \frac{M_{\rm p}}{0.05 M_\odot} \right) 
       \left( \frac{\eta_d}{0.1} \right) 
         \left( \frac{\eta_j}{0.1} \right)^{} 
     \\
     & \times
       \left( \frac{M_I}{0.1 M_\odot} \right)^{-1} 
         \left( \frac{v_j}{20 v_I} \right)^{2} .
     \end{split}
\label{eq:FjI}
\end{equation}

Equation (\ref{eq:FjI}) is a crude estimate of a plausible scenario where a massive planet or a brown dwarf leads to the launching of jets. Equation (\ref{eq:FjI}) crudely indicates that the jets can shape the nebular gas along their propagation trajectory. We suggest that the jets that resulted from the merger of the sub-stellar object with the main sequence companion led to the launching of two pairs of jets along almost the same direction. These pairs inflated the two low-brightness lobes, one with the two rims on the south of the inner structure of NGC 1514, and one to the northeast. There were two closely spaced in time jet-launching episodes, as evident from the rims (panel a of Figure \ref{fig:NGC1514iamges}). In addition, shortly before or after these two jet launching episode following the sub-stellar merger, there was also a jet launching episode from mass that the main sequence accreted from the AGB star along the main axis of this PN, as evident from the two ears along the main axis of NGC 1514; the axis marked by the double-ended red arrow on panel c of Figure \ref{fig:NGC1514iamges}.
The simulations of the messy morphology of the inner structure of NGC 1514 are outside the scope of the present study. We turn to simulate the formation of jet-shaped bipolar double-ring structures.

\section{The outer rings by jets}
\label{sec:outrings}

\subsection{Numerical setup}
\label{subsec:Numerical}

We conduct three-dimensional (3D) hydrodynamical simulations using version 4.8 of the {\sc flash} code \citep{Fryxelletal2000}, employing the unsplit hydro solver \citep{LeeDeane2009}. FLASH is a modular, adaptive-mesh refinement (AMR) hydrodynamics and magnetohydrodynamics code. We do not include gravity, as the velocities of the gas are much larger than the escape velocity from the regions we simulate. In all simulations we used a Courant–Friedrichs–Lewy number of 0.3. 
For the Riemann solver we adopted the HLLC method \citep{Toro1994}, together with artificial viscosity to stabilize shocks.
The computational domain is a Cartesian grid $(x,y,z)$ with outflow boundary conditions at all boundaries. The plane $z=0$ is the stellar and binary equatorial plane. Our simulations cover both sides of the equatorial plane. The total sizes of the grid in each direction are $L_x = Ly=4 \times 10^{17} \cm $, and  $L_z = 6 \times 10^{17} \cm$. 
We utilize a 3D AMR grid with eight refinement levels, resulting in the best resolution with a cell size of $2^{-10}L$, where $L$ is the grid size. The numerical code splits cells to reach higher resolution in zones where there are large gradients in one or more flow quantities. For the parameters we use, the grid cells are boxes, with the best resolution of $(\Delta_x, \Delta_y, \Delta_z)=2^{-10}(L_x,L_y,L_z)=( 3.9 \times 10^{14} \cm, 3.9 \times 10^{14}\cm, 5.86 \times 10^{14}\cm)$.
We simulated two cases with a lower level of resolution, where the best resolution has twice as large cells. We still obtain the rings, but they are smeared somewhat, as expected. We therefore find that our simulations converge in the highest resolution we use. With our computer resources, we could not simulate a higher level of resolution.

At time $t=0$, we introduce a dense spherical shell around the center, as we applied in \cite{Akashi2017Barrel}, where a schematic figure can be found. We launch two central opposite jets perpendicular to the $z=0$ plane that interact with the shell. We aim to show that two opposite jets interacting with a dense shell can form two rings with morphology as observed in PN NGC 1514. 

We set the initial radial velocity of the shell to be $v_s = 10 \km \s^{-1}$; this is a much slower velocity than the jets' velocity, and its exact value is not of great significance to the goals of this study. 
We simulated cases with two different shells. One shell occupies the radial region $10^{17} \cm < r < 1.1 \times 10^{17} \cm$, with the density profile 
$\rho_s = 1.6 \times 10^{-19} (r/10^{17} \cm)^{-2} \g \cm^{-3}$, 
resulting in a total shell mass of $M_s = 0.1 M_\odot$. A mass loss episode with a mass loss rate of $\dot M_s = 3.2 \times 10^{-4} M_\odot \yr^{-1}$ that lasted for $\Delta t_s = 315 \yr$ formed this shell. The other shell we simulate occupies the radial region $10^{17} \cm < r < 1.2 \times 10^{17} \cm$, with the density profile 
$\rho_s = 8 \times 10^{-20} (r/10^{17} \cm)^{-2} \g \cm^{-3}$, 
resulting $M_s = 0.1 M_\odot$, $\dot M_s = 1.6 \times 10^{-4} M_\odot \yr^{-1}$, and $\Delta t_s = 630 \yr$.  We tested the evolution of the shell when we do not inject jets. The shell expands because of the initial radial velocity, and widens because of internal pressure. What is important to the present study is that for the 220 years of the simulations we present here, the shell maintains its smooth and spherical structure. The instabilities we will report on are real and do not result from an unstable shell expansion.


The density in the inner and outer zones of the dense shell is of a slow, $v_{\rm wind} = v_s = 10 \km \s^{-1}$, spherically symmetric, low-mass loss rate wind, $\dot M_{\rm wind} = 1.5 \times 10^{-6} M_\odot \yr^{-1}$. The density profile in these zones is  
$\rho_{\rm wind}(r) = 7.5 \times 10^{-22} (r/10^{17} \cm)^{-2}$.

 We numerically initialize the jets by defining two oppositely directed conical regions with a prescribed half-opening angle and injection radius. Within these cones, at each timestep during the jet active phase, the gas cells are assigned the jet velocity, density, and temperature that correspond to the desired jet properties (mass-loss rate, kinetic, and thermal power). The injection is maintained for the specified jet active time. This prescription ensures a continuous, steady injection of bipolar jets with well-defined parameters into the computational domain. In the present simulations,  we inject two oppositely directed jets from a central region within a radius of $10^{16} \cm$, aligned along the $z$-axis and confined within a half-opening angle of $\alpha_j = 45^\circ$. We terminate the jet launching at $t = 48 \yrs$. The jets are injected into the grid with a velocity of  
$v_j = 800 \km \s^{-1}$,  
slightly above the escape velocity from a main-sequence star. The combined mass-loss rate into both jets is  
$\dot M_{\rm 2jets} = 1.1 \times 10^{-4} M_\odot \yr^{-1}$.

At the beginning of the simulation, all components—the slow wind, the dense shell, and the jets—are set to a temperature of $10^4 \K$. The specific initial temperature of the jets has a small impact on the dynamics, provided they are highly supersonic, as they cool quickly through adiabatic expansion.

In our simulations, we adopt an ideal gas equation of state with a uniform composition. Ionization is not included, and the gas is treated as neutral.

In our simulations, we adopt an ideal-gas equation of state with a constant adiabatic index $\gamma$, and a uniform composition of a neutral gas. We do not follow the ionization of the gas. We follow the global structure for which ionization will not have a large effect. As well, following ionization requires including the ionizing radiation of the central star, which is beyond the present scope. To represent different thermodynamic conditions, we use two values: $\gamma = 5/3$ and $\gamma = 1.05$. The value $\gamma = 5/3$ corresponds to a monoatomic ideal gas and is appropriate when radiative losses are negligible (often referred to as "adiabatic evolution", even though entropy changes at shocks). To account for the effects of radiative cooling in an approximate manner, we also perform simulations with a significantly lower adiabatic index, $\gamma = 1.05$. This low value mimics the cooling effect due to photon diffusion, allowing the gas to lose energy more efficiently and thereby approximating nearly isothermal behavior. This method has been commonly employed in astrophysical hydrodynamic simulations where explicit radiative transfer is computationally prohibitive (see our discussion in \citealt{AkashiSoker2013}).

In addition to $\gamma<5/3$ simulations, we also perform simulations with $\gamma=5/3$ but with radiative cooling of an optically thin plasma. In these simulations, we implement radiative cooling in FLASH by adding a tabulated cooling function from \cite{SutherlandDopita1993}. The code interpolates the table during runtime to obtain the cooling function as a function of temperature. FLASH treats this cooling term in an operator-split manner: after the hydrodynamical update, the internal energy is updated separately according to the cooling rate. This procedure, standard in FLASH,  ensures numerical stability even for short cooling timescales by using sub-cycling if needed. The method is first-order accurate in time with respect to the cooling source term, as is customary in operator-split implementations. 

We aim to show the formation of the morphology of two rings and discuss the conditions for their formation by two opposite jets. The above set of parameters will serve us to scale to other cases, when the rings might be formed much closer to the star. For example, in both cases of the shell, the shell ejection episode ended at a time of $t_{\rm end,s}=-3170 \yr$ (3170 years before we start the simulation). However, our results, under the correct usage of radiative cooling that we discuss later, can be applied to the interaction of jets with a shell that is, say, 100 times smaller and has an expanding velocity of, say, $50 \km \s^{-1}$, which is still much slower than the jets' velocity. That shell will result from a much higher mass loss rate. In the latter case, the jet launching episode starts only 6 years after the shell formation episode. The primary difference between various sizes and mass loss rates lies in the role of radiative cooling, specifically whether it is significant or not during the jet-shell interaction. We will show that our setting forms two rings when radiative cooling is small.

\subsection{Results}
\label{subsec:results}

We present the results of our hydrodynamical simulations, designed to investigate the formation of bipolar rings through the interaction of jets with a dense spherical shell. The simulations aim to interpret the morphology of planetary nebulae, such as NGC 1514 (e.g., \citealt{Ressleretal2025}) and Abell 14 (for the rings see, e.g., \citealt{Akras2016}, \citealt{Akras2020}). We present many maps in the meridional $y=0$ plane, and at three representative times: $t = 60~\mathrm{yr}$, $t = 140~\mathrm{yr}$, and $t = 220~\mathrm{yr}$. These times were selected to highlight key stages in the ring formation process, from shortly after we terminate the jets to several times the jet activity period of $48 \yr$. Again, our results can be scaled to much shorter interactions, provided that the mass loss rates are appropriately scaled for the dense shell and the jets.

\subsubsection{Density evolution: ring formation}

\begin{figure*}[htb]
\centering
\includegraphics[trim=0.0cm 6.5cm 0.0cm 0cm,width=1.00\textwidth]{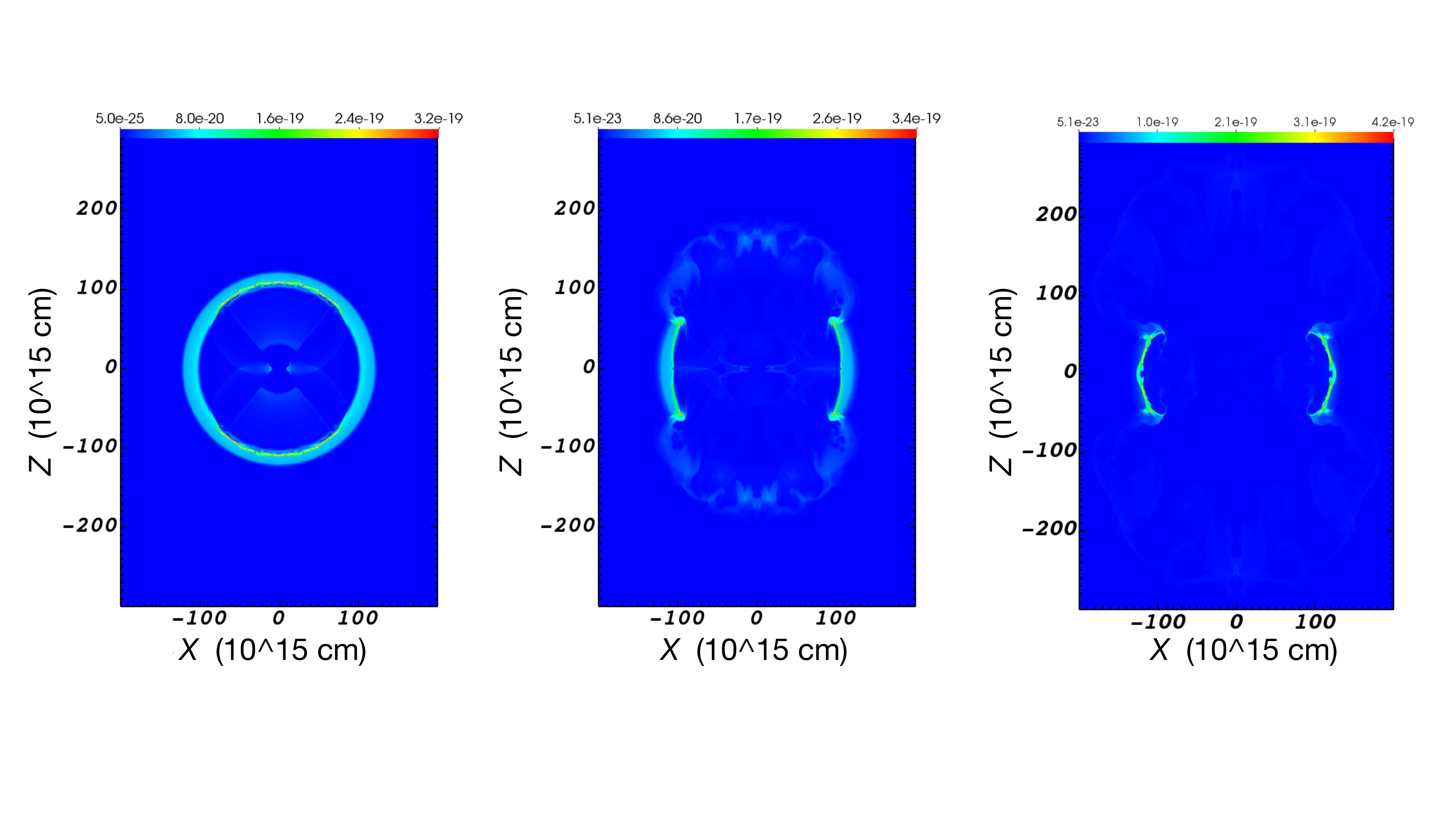}
\caption{Density maps in the $y=0$ meridional plane at three times: $t = 60~\yr$ (left), $t = 140 \yr$ (middle), and $t = 220~\yr$ (right). The density is given in units of $\mathrm{g~cm^{-3}}$, as indicated by the color bar at the top; note that the scale of the color bar varies across the different panels. In this figure, the slow spectral shell is the wide one and $\gamma = 5/3$. In the meridional plane, two arcs evolve, with dense regions at their edges (note the red color at the edges). These arcs are the cross-section of a barrel-like structure in 3D, and the dense edges of the arcs form two opposite rings in 3D. }
\label{fig:densxz_gama167}
\end{figure*}

Figure~\ref{fig:densxz_gama167} shows the density maps in the meridional plane and at three times of the simulations with the initial wide dense shell and $\gamma=5/3$ with no radiative cooling. At early times ($t = 60~\mathrm{yr}$), the jets are still inside the dense shell and compressing material along the polar directions. After the jets break out from the dense shell, they inflate low-density polar bubbles and leave behind a dense barrel-like structure. The cross-section of the barrel-shaped 3D structure on the meridional plane is two arcs, as the middle panel of Figure~\ref{fig:densxz_gama167} shows at $t = 140~\mathrm{yr}$. The edges of these arcs are denser (contain a small red zone) than the rest of the arcs, and in 3D form a structure of two rings. This structure of two rings becomes more prominent at a later time of $t = 220~\mathrm{yr}$, which we present in the right panel of Figure~\ref{fig:densxz_gama167}. 

\begin{figure*}[htb]
\centering
\includegraphics[trim=0.0cm 6.5cm 0.0cm 0cm,width=1.00\textwidth]{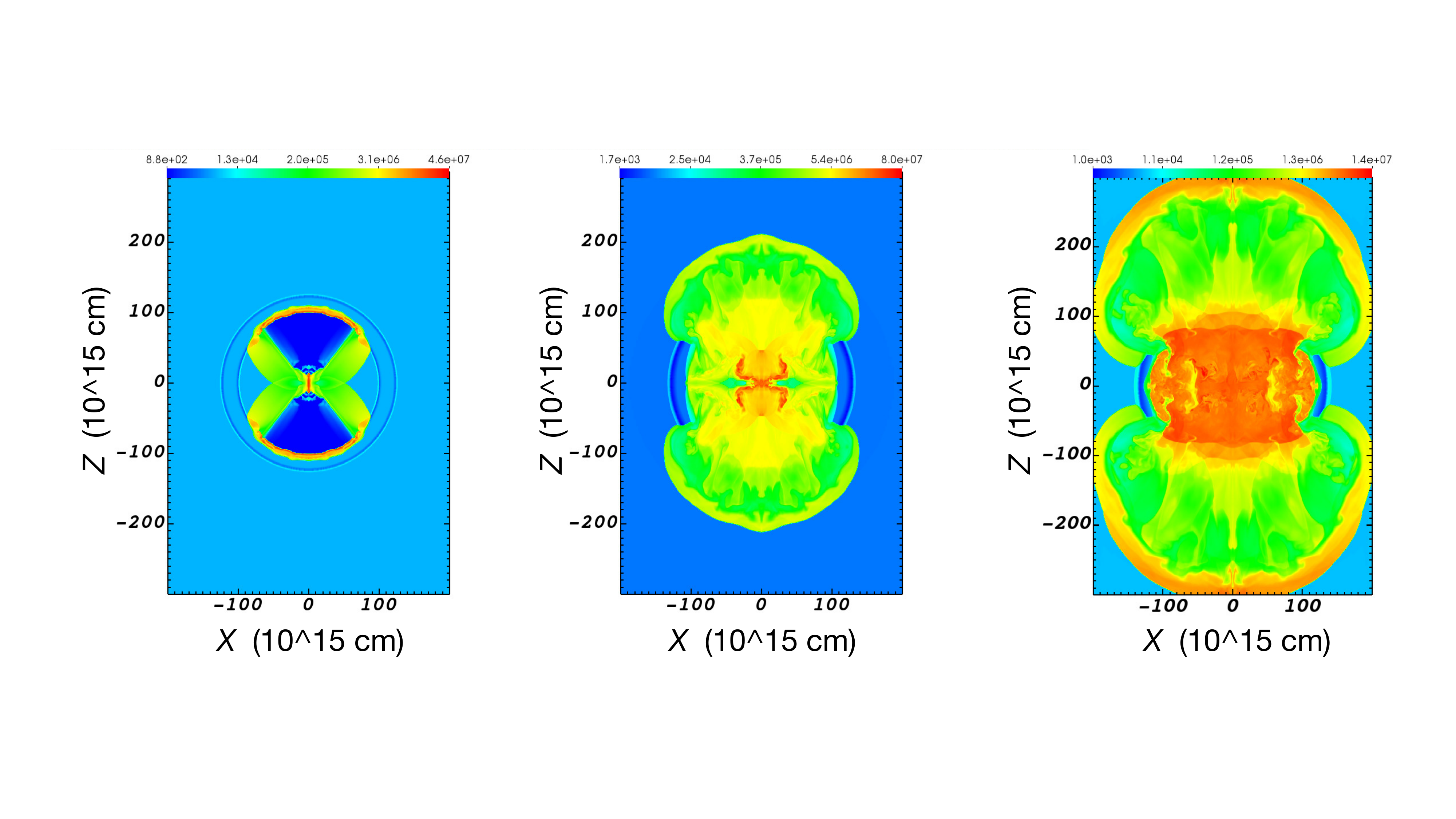}
\caption{Temperature maps in the same plane, the same times, and the same simulation as in Figure~\ref{fig:densxz_gama167}. Temperatures are expressed in Kelvin, as indicated by the color of the bar; note the varying upper values in the different panels ($4.6 \times 10^7 \K$, $8.0 \times 10^7 \K$, and $1.4 \times 10^7 \K$. Note the hot bubble developing in the center,  and the filamentary temperature structure at late times that result from Rayleigh-Taylor instabilities. } 
\label{fig:temp_gama167}
\end{figure*}

An interesting feature that appears is the compression of the surviving dense shell segments, which appear as two thin arcs in the density maps in the meridional plane; they are thinner than the initial shell. Namely, despite the jets having a half-opening angle of $\alpha_j = 45^\circ$, they also compress the shell near the equatorial plane. The reason is that the shocked jet material forms a hot bubble in the center, which exerts pressure from within on the entire dense shell. In Figure~\ref{fig:temp_gama167} we present the temperature maps in the same plane and at the same times as in Figure~\ref{fig:densxz_gama167}. While at earlier times the large regions with high temperatures are the jet-shell interaction regions, at later times the high temperature is of the hot central bubble. This forms a high-pressure central bubble that compresses the remnant of the shell, now a barrel-like structure, explaining the thinning of the two arcs seen in Figure~\ref{fig:densxz_gama167}.  

\begin{figure*}[htb]
\centering
\includegraphics[trim=0.0cm 8.0cm 0.0cm 0cm,width=1.00\textwidth]{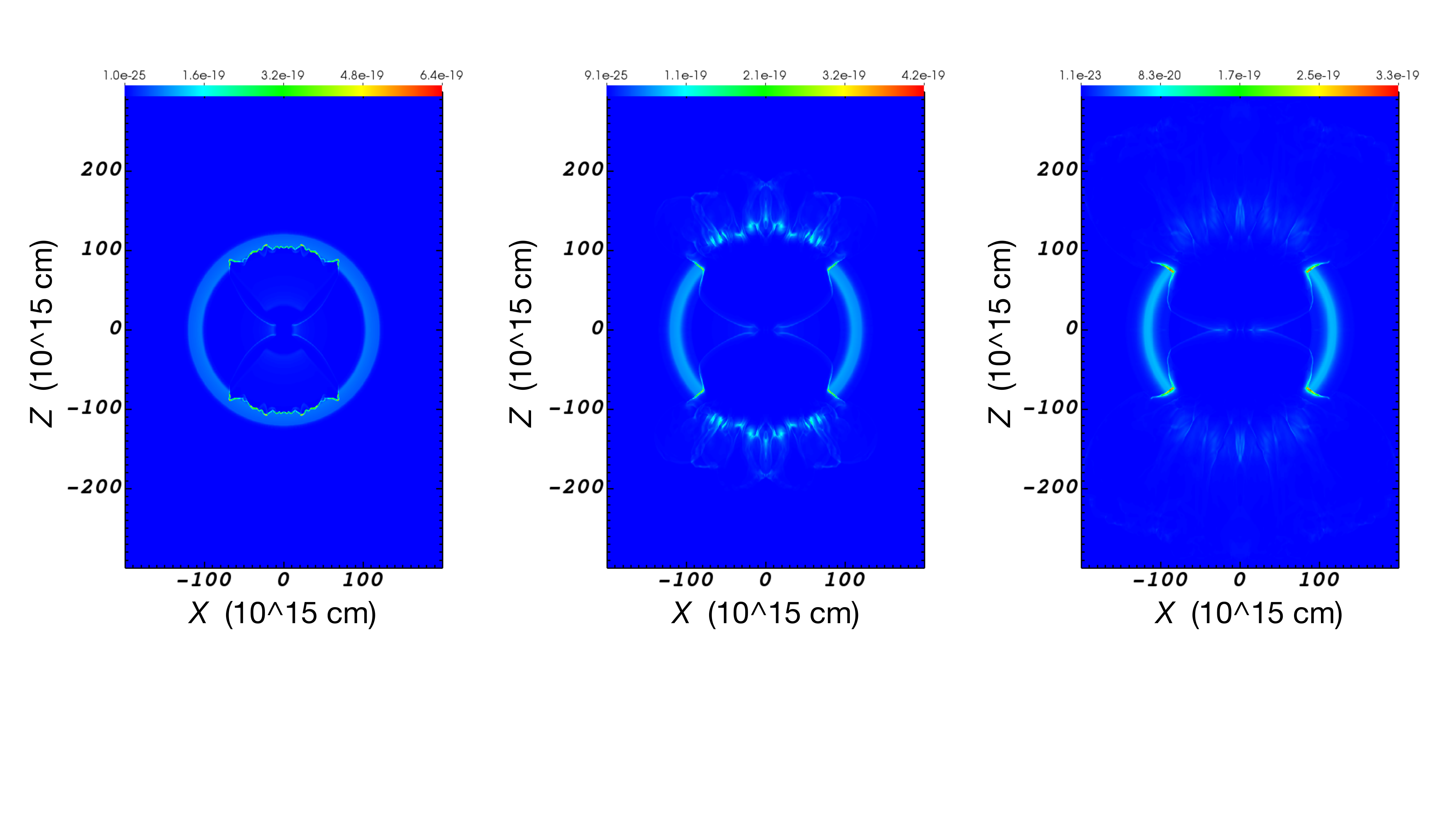}
\caption{Similar to Figure~\ref{fig:densxz_gama167} but for a simulation with $\gamma=1.05$. }
\label{fig:densxz_gama105}
\end{figure*}

\begin{figure*}[htb]
\centering
\includegraphics[trim=0.0cm 5.0cm 0.0cm 0cm,width=1.00\textwidth]{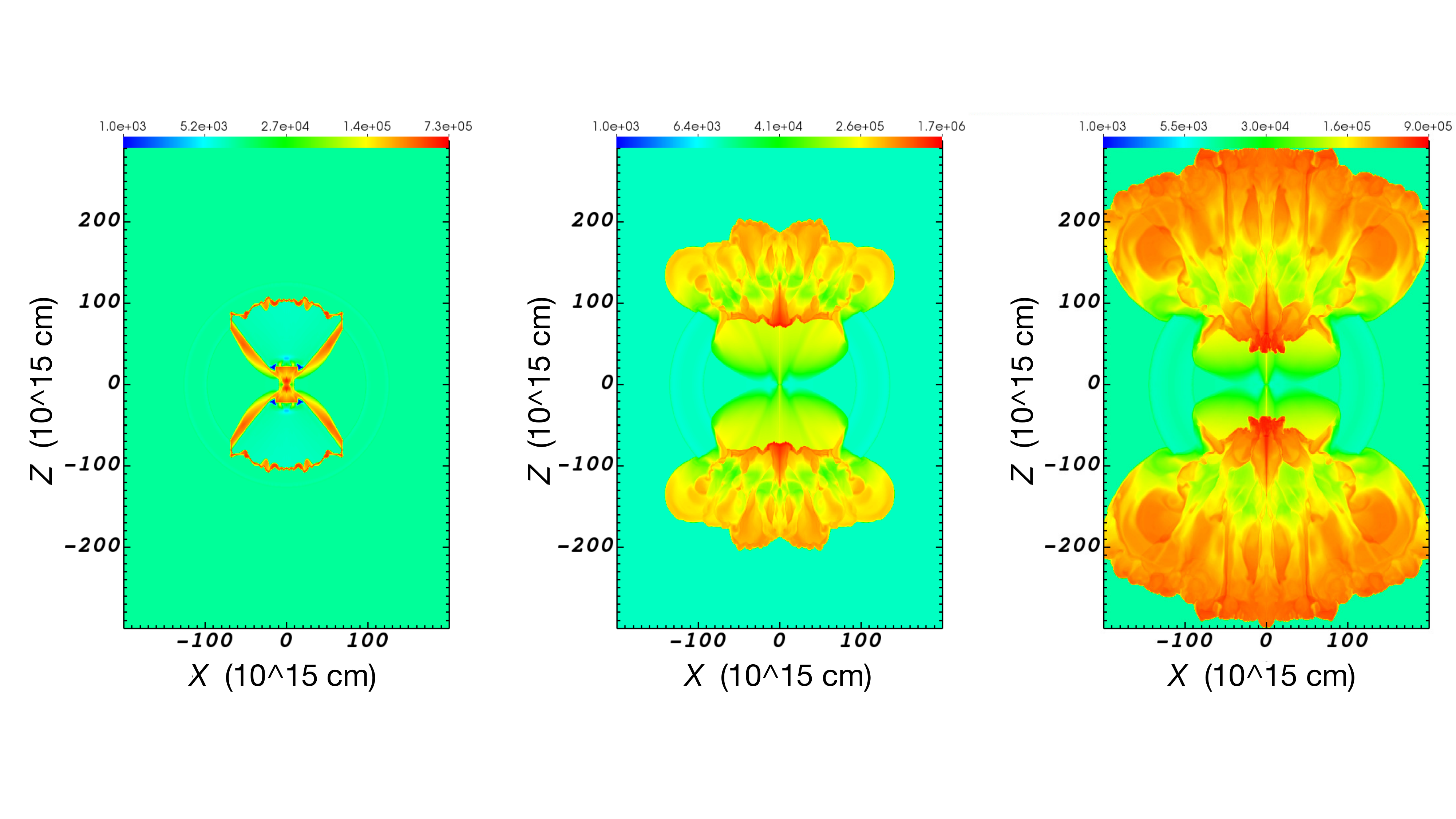}
\caption{Similar to Figure~\ref{fig:temp_gama167} but for a simulation with $\gamma=1.05$. Top temperatures in the color bars (deep red), from left to right, are $7.3 \times 10^5 \K$, $1.7 \times 10^6 \K$, and $9.0 \times 10^5 \K$. }
\label{fig:temp_gama105}
\end{figure*}

In Figures~\ref{fig:densxz_gama105} and \ref{fig:temp_gama105} we present the density and temperature maps in the meridional plane of the simulation with an adiabatic index of $\gamma=1.05$, at the same times as in Figures~\ref{fig:densxz_gama167} and \ref{fig:temp_gama167}. Apart from the value of $\gamma$, the parameters of the two simulations are identical. The significantly lower value of $\gamma$ mimics a case where radiative cooling is substantial; therefore, the thermal energy in the shocked gas is substantially lower than in the non-radiative case. The interaction is momentum-driven rather than energy-driven (for studies on momentum- and energy-conserving flows, see, e.g., \citealt{Garcia2000, Mellema2001, Mellema2003}). 

Both simulations produce a barrel-shaped structure from the remnant of the dense shell, featuring two dense rings at the edges of the remnant. Both simulations inflate two bipolar lobes, but of different structures.  
In the two simulations, the jet-shell interaction in the polar directions is prone to Rayleigh-Taylor instabilities. This is evident from the clumpy polar shell in the middle panel of Figure~\ref{fig:densxz_gama167} and the two right panels of Figure~\ref{fig:densxz_gama105}. 
The temperature maps at late times (right panels in Figures \ref{fig:temp_gama167} and \ref{fig:temp_gama105}), present filamentary structures inside the bubbles that result from Rayleigh-Taylor instabilities. We mapped (not shown) Rayleigh-Taylor unstable zones, and they also exhibit a filamentary structure, in addition to the instability at the front of the inflated bubble. These filaments are not along the grid axis and are curved. Therefore, these instabilities are Rayleigh-Taylor physical instabilities and not numerical instabilities.

The main differences between the non-radiative and the $\gamma=1.05$ simulations, as shown in the density and temperature maps, are as follows. 
(1) While the non-radiative interaction inflates two large polar bubbles expanding in all directions, as seen in the last temperature maps, the $\gamma=1.05$ accelerates material in the polar directions, as expected from a momentum-conserving flow. 
(2) The non-radiative simulation disperses the polar shell gas at early times, as no polar gas is seen in the right panel of  Figure~\ref{fig:densxz_gama167}. Dense clumps are seen in the right panel of Figure~\ref{fig:densxz_gama105}. 
 (3) At late times, the non-radiative simulation develops a hot central bubble, which the $\gamma=1.05$ does not. 
(4) The hot central bubble in thenon-radiative simulation compresses the remnant of the dense shell. This does not occur in the $\gamma=1.05$ simulation. 

Although both simulations produce a pair of bipolar rings, in Section \ref{subsubsec:Projection} we demonstrate that the non-radiative one more closely matches the observed double-ring morphology of PN NGC 1514. 

\subsubsection{The flow structure}
\label{subsubsec:Velocity}

\begin{figure*}[htb]
\centering
\includegraphics[trim=0.0cm 4.5cm 0.0cm 0cm,width=1.00\textwidth]{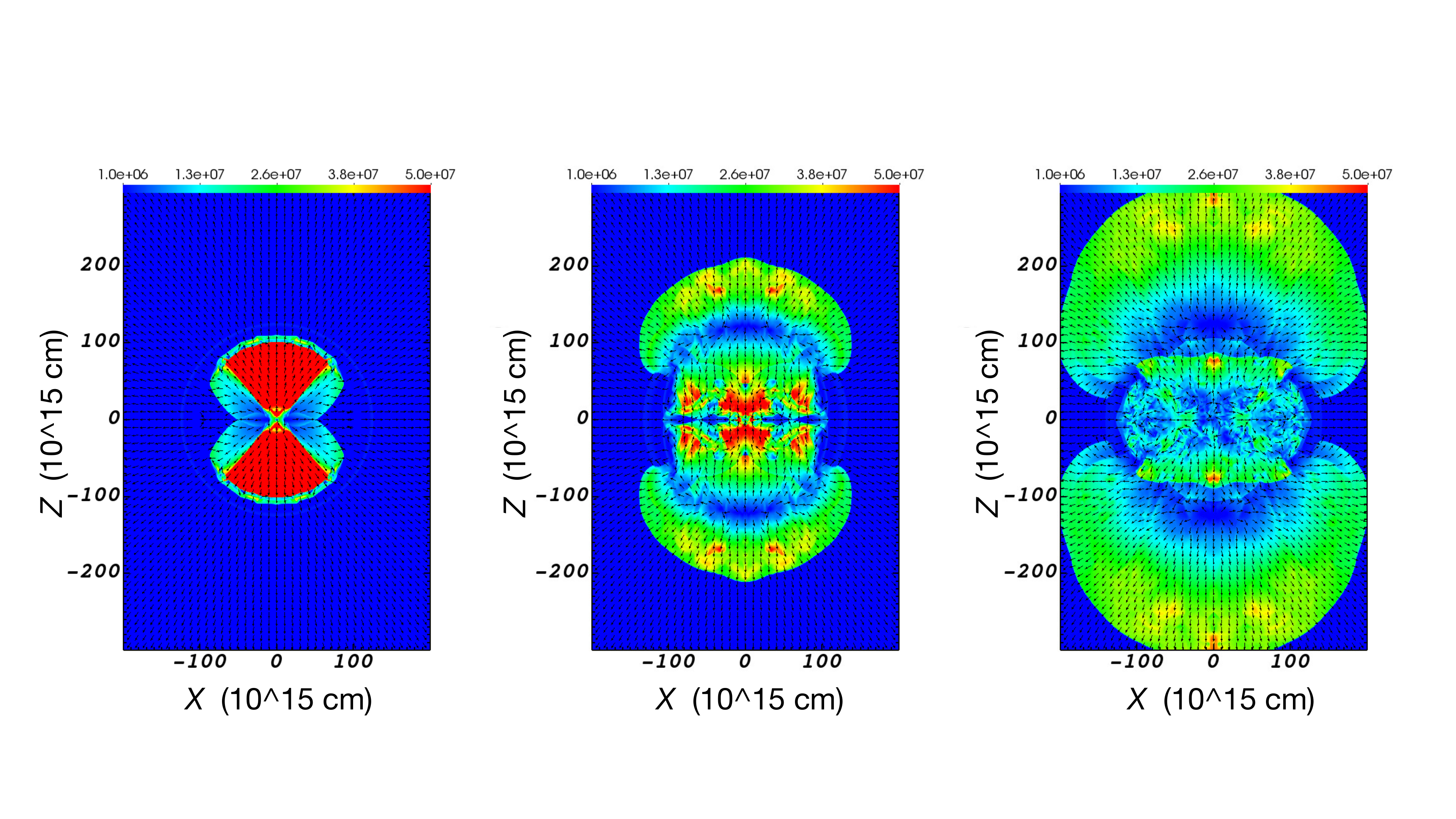}
\caption{Velocity maps in the same meridional plane, for the same three times, and the same simulation as in Figure \ref{fig:densxz_gama167} and \ref{fig:temp_gama167}. Arrows show the flow direction and colors the velocity magnitude in $\cm \s^{-1}$ according to the color bar from $10^6 \cm^{-1}  = 10 \km \s^{-1}$ (deep blue) to $ 5 \times 10^6 \cm^{-1}  = 500 \km \s^{-1}$ (deep red).  
}
\label{fig:vel_gama167}
\end{figure*}

\begin{figure*}[htb]
\centering
\includegraphics[trim=0.0cm 8.5cm 0.0cm 0cm,width=1.00\textwidth]{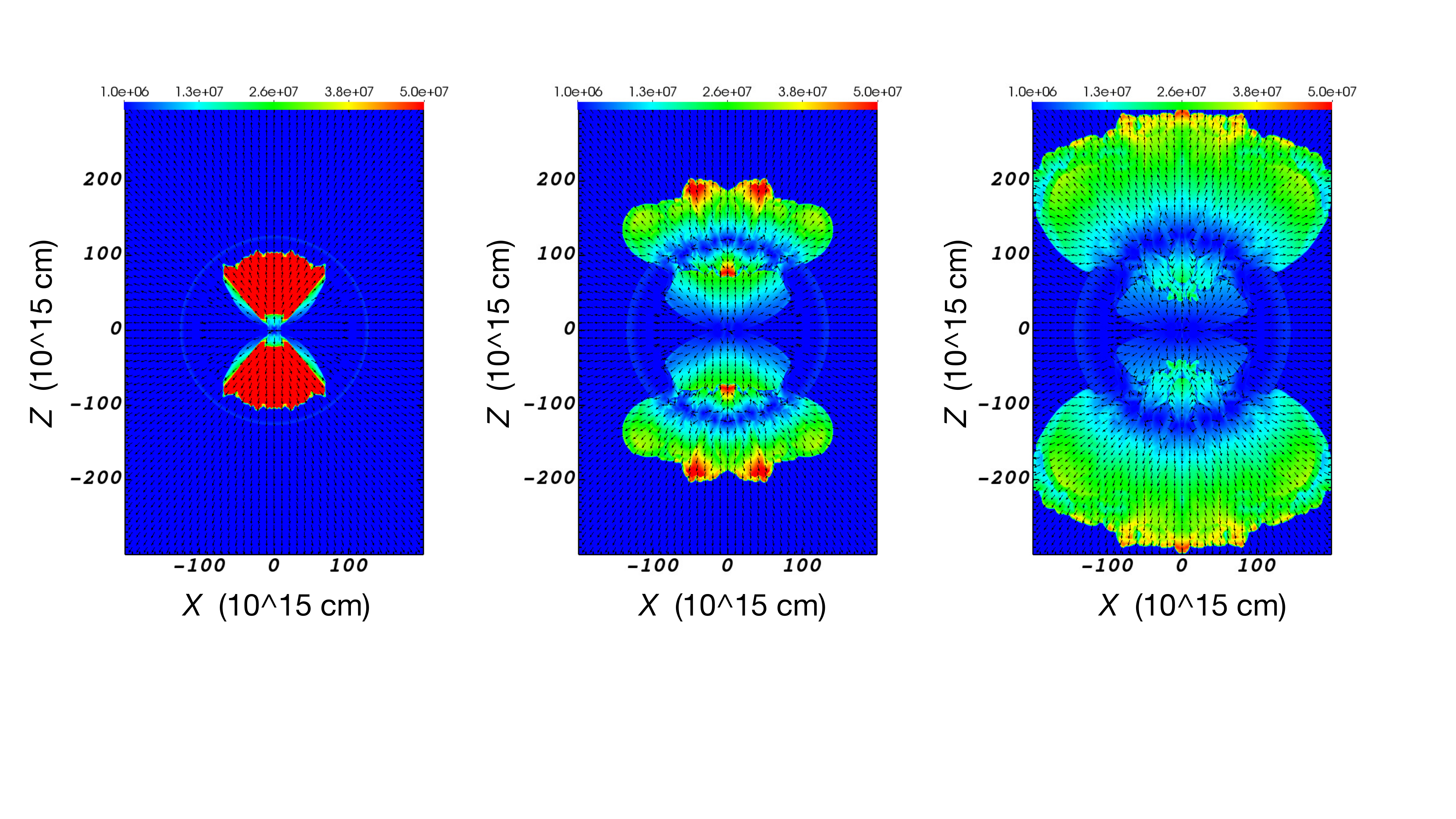}
\caption{Similar to Figure~\ref{fig:vel_gama167} but for $\gamma = 1.05$.}
\label{fig:vel_gama105}
\end{figure*}

To better reveal the flow structure, we present in Figures~\ref{fig:vel_gama167} and \ref{fig:vel_gama105} the velocity maps for the non-radiative and $\gamma=1.05$ simulations at the same time and plane as in Figures~\ref{fig:densxz_gama167} and \ref{fig:densxz_gama105}, respectively; the arrow present the direction of the flow and the color the velocity magnitude, from $10 \km \s^{-1}$ (deep blue) to $500 \km \s^{-1}$ (deep red). At early times, the fastest flows (green/yellow/red) are bipolar, corresponding to the launching directions of the jets. At later times, the two simulations differ. 

As expected, most of the fast-moving material in the momentum-conserving simulation of $\gamma=1.05$ (Figure~\ref{fig:vel_gama105}) more or less maintains its bipolar outflow. In the non-radiative simulation, on the other hand, fast vortices develop around the center in the hot central bubble (Figure~\ref{fig:vel_gama167}). These vortices play a role in compressing the barrel-shaped remnant of the dense shell (Figure~\ref{fig:densxz_gama167}). Additionally, the flow of the bipolar lobes in the non-radiative simulation, which is outside the rings (i.e., at larger distances from the center), exhibits lateral expansion, resulting in significant velocities at intermediate latitudes. 

Both simulations exhibit instabilities at the front of the jets, resulting in a clumpy bipolar outflow, as illustrated in all figures (density, temperature, and velocity). These are Raleigh-Taylor instabilities that develop as the low-density shocked jets' material accelerates the dense shell.  The high-pressure gas escapes through the instability fingers, releasing the pressure inside, leaving behind slower-moving clumps along the polar direction, more prominent in the momentum-conserving simulation of $\gamma=1.05$.  

\subsubsection{Mimicking observational images}
\label{subsubsec:Projection}

To facilitate comparison with observations, we generate column-density maps projected onto the plane of the sky by integrating the density along a line of sight. We present results for different inclination angles $i$, which is the angle between the simulation's symmetry axis (the $z$-axis) and the line of sight. We integrate the density, not the density squared, as we aim to reproduce the IR emission from dust. Specifically, \cite{Ressleretal2025} find that the IR emission of the rings is thermal emission from very small grains.  

\begin{figure*} 
\centering
\includegraphics[trim=0.0cm 16.2cm 0.0cm 0cm,width=1.08\textwidth]{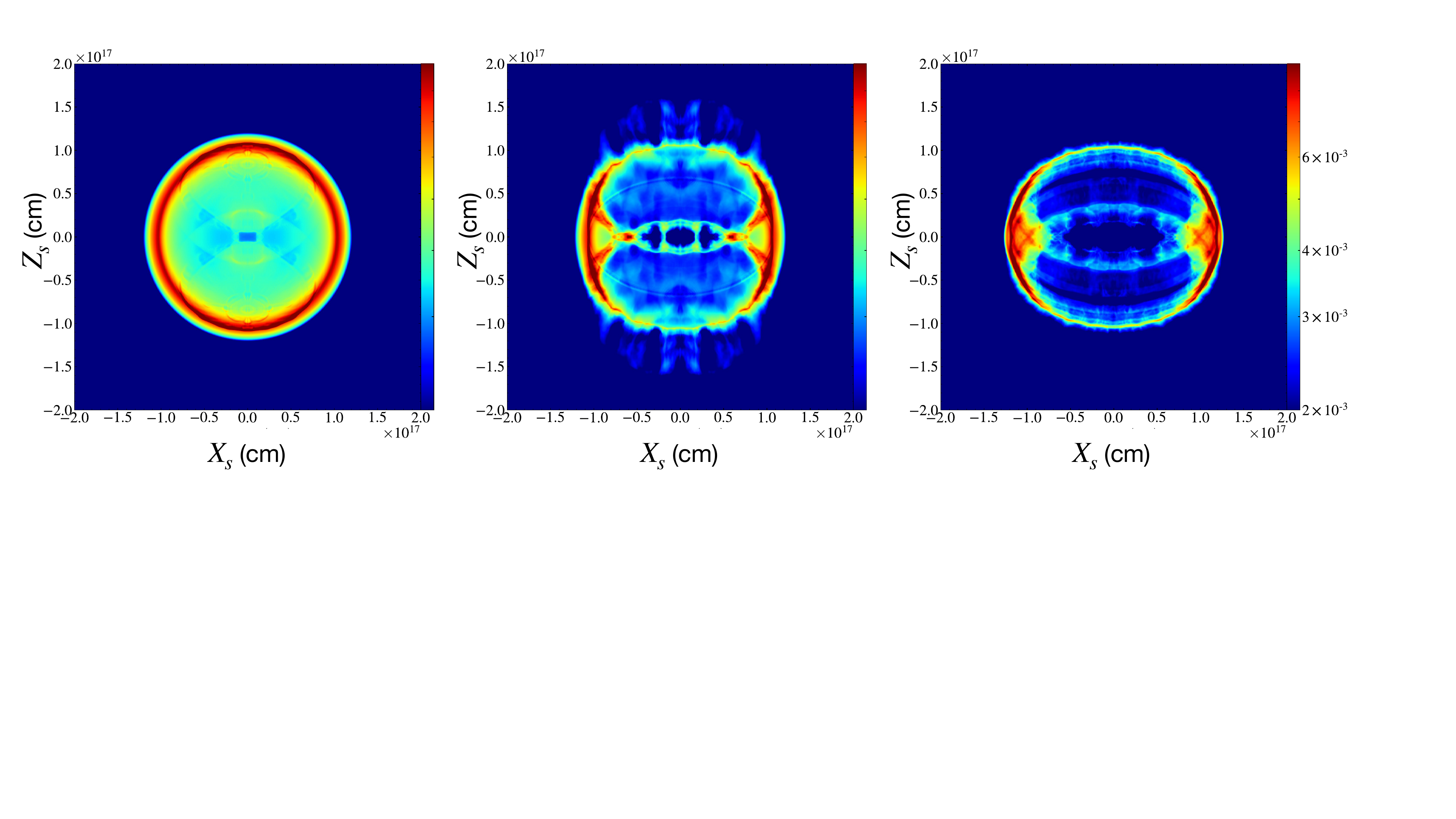}
\caption{Column-density maps obtained by integrating $\int \rho\, \mathrm{d}l$ along a line of sight inclined by $i=50^\circ$, which is the angle between the line of sight and the $z$-axis of the simulation, for the same simulation, $\gamma = 5/3$, and same times as in Figures \ref{fig:densxz_gama167}, \ref{fig:temp_gama167}, and \ref{fig:vel_gama167}.  The coordinates on the plane of the sky are $X_S$, which is parallel to the numerical X-axis, and $Z_S$, which is in the YZ plane of the numerical grid. 
The two bipolar rings appear in the last panel as enhanced arcs at intermediate latitudes. Units are according to the color bar from $0.002 \g \cm^{-2}$ in deep blue to $0.009 \g \cm^{-2}$ in deep red. }
\label{fig:proj_gama167}, 
\end{figure*}

\begin{figure*} 
\centering
\includegraphics[trim=0.0cm 11.2cm 0.0cm 2.0cm,width=1.08\textwidth]{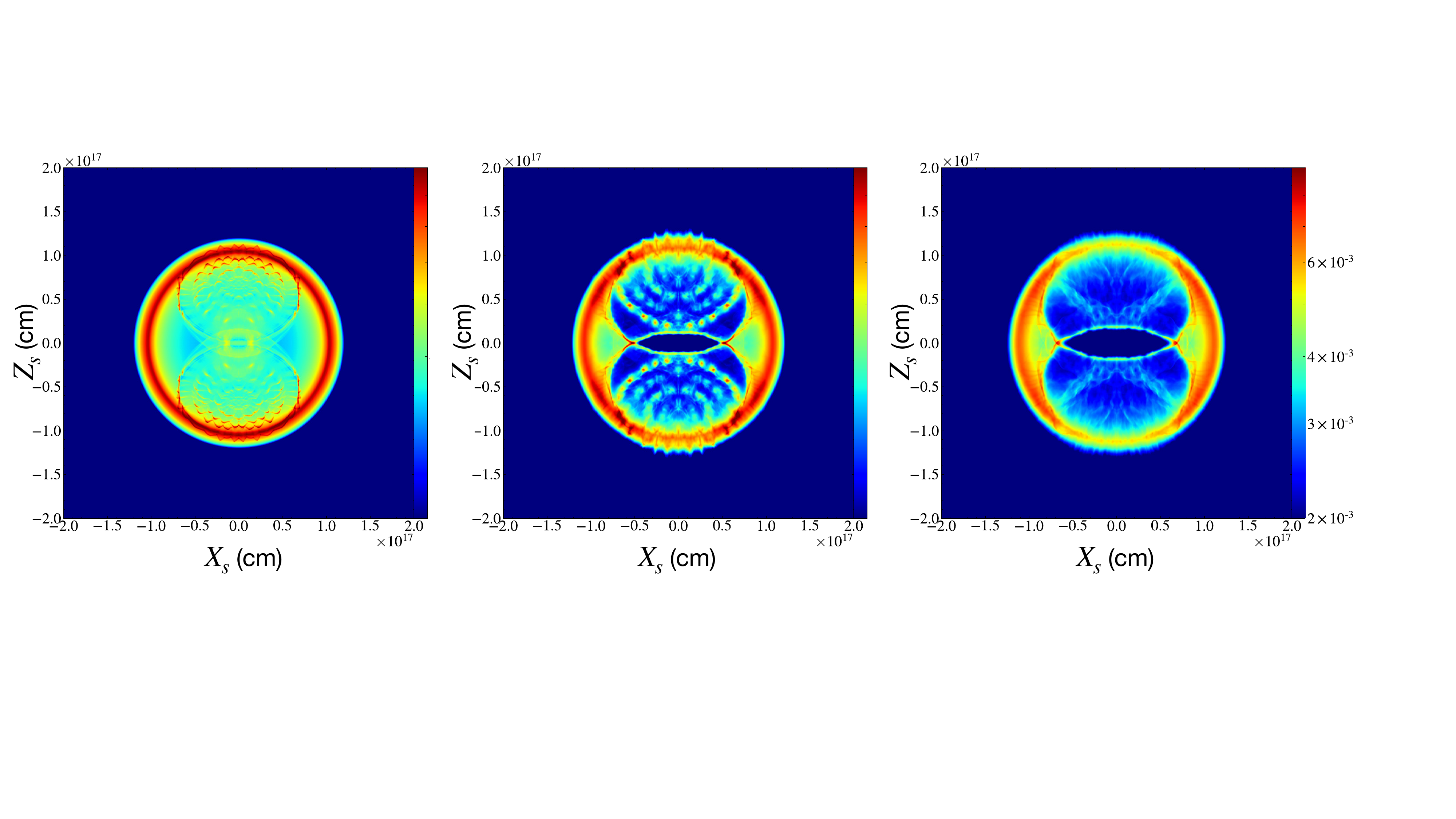}
\caption{Similar to Figure~\ref{fig:proj_gama167} but for the simulation with $\gamma = 1.05$. In this case, the last panel shows very thin rings, much thinner than those observed in NGC 1514. }
\label{fig:proj_gama105}
\end{figure*}

In Figures~\ref{fig:proj_gama167} we present the column density, in units of $\g \cm^{-2}$ according to the color bar, at three times as in Figure \ref{fig:densxz_gama167}, and for $i=50^\circ$. The coordinates are those on the plane of the sky, where $X_S$ is parallel to the numerical $X$-axis and $Z_S$ is in the YZ plane of the numerical grid. In Figure~\ref{fig:proj_gama105} we present the column density for the simulation with $\gamma=1.05$, at the same inclination angle and times as in Figure~\ref{fig:proj_gama167}. Both simulations produce a bipolar double-ring system. However, while the non-radiative simulation $\gamma=5/3$ produces prominent thick rings, the $\gamma=1.05$ simulation produces thin rings that do not resemble the rings of NGC 1514.  

\begin{figure} 
\centering
\includegraphics[trim=4.7cm 0.0cm 0.0cm 0cm,width=0.68\textwidth]{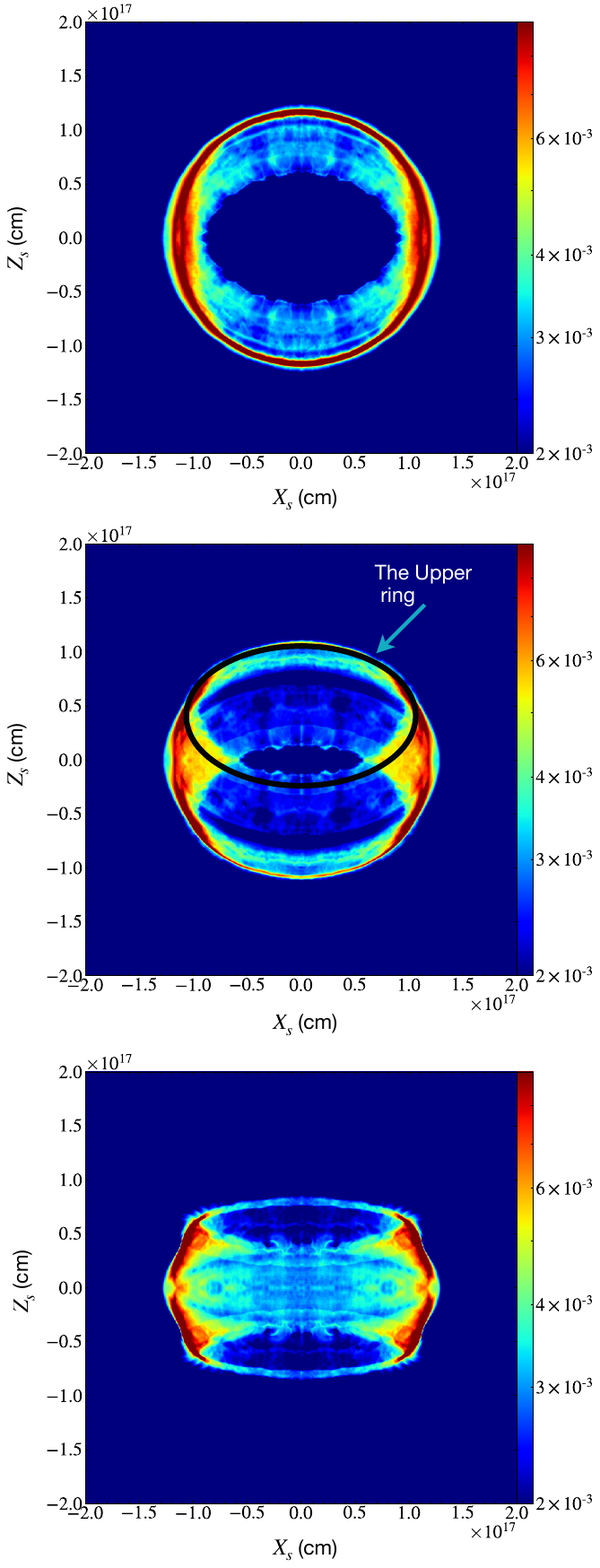}
\caption{Column density maps for the run with adiabatic index $\gamma = 5/3$ and a thin shell, shown at $t = 220 \yr$. From top (close to pole-on) to bottom (close to edge-on), the inclination angles are $i=25^\circ$, $i=50^\circ$, and $i=75^\circ$, respectively.  }
\label{fig:thinshell_differentangles}
\end{figure}

In Figure \ref{fig:thinshell_differentangles} we present the column density of the non-radiative case $\gamma=5/3$, where the initial AGB dense shell is thin, namely, half the initial thickness of the results we presented in previous figures. We present the column density map at the last time, $t=220 \yr$, but at three inclination angles, $i=25^\circ$ (upper panel; close to being pole-on), $i=50^\circ$ (middle panel), and $i=75^\circ$ (lower panel; close to being edge-on). The thin shell simulation also produces the bipolar double structure as in NGC 1514 for an inclination angle of $i \simeq 50 ^\circ$; we mark one ring with a black ellipse in the middle panel. The large inclination angle, almost edge-on, presents two rings, on the upper and lower parts of the simulated nebula. These two rings resemble the two rings of PN Abell 14 (e.g., \citealt{Akrasetal2016} for an image of Abell 14), although in Abell 14, there is a structure outside the two rings. 

\begin{figure} 
\centering
\includegraphics[trim=0.cm 0.0cm 0.0cm 0cm,width=0.4\textwidth]{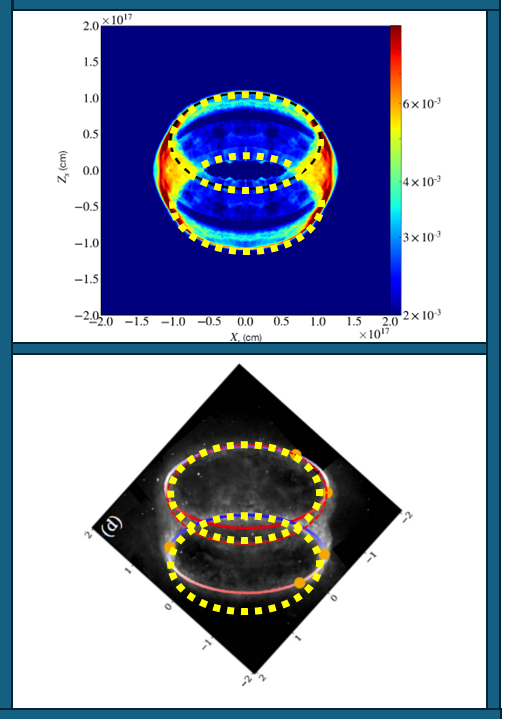}
\caption{A comparison of the column density of the non-radiative simulation with the thin shell at $i=50^\circ$ from the middle panel of Figure \ref{fig:thinshell_differentangles} to an IR image of NGC 1514 (panel d of Figure \ref{fig:NGC1514iamges} adapted from \citealt{Ressleretal2025}, and rotated). We mark the rings produced by the simulated jets with yellow dotted ellipses and copied them to the IR image of NGC 1514 in the lower panel. The upper simulated ring fits the observed one, while there is a small departure in the fitting of the lower ring, resulting from departure from a mirror symmetry of NGC 1514 (Section \ref{sec:NGC1514}).}  
\label{fig:Comparison}
\end{figure}

Figure \ref{fig:Comparison} compares the column density of the non-radiative simulation with a thin shell at $i=50^\circ$ that we present in the middle panel of Figure \ref{fig:thinshell_differentangles} with an IR  observation of NGC 1514 (lower panel). We mark the two rings that we identify in this simulation by dashed-yellow ellipses (marked by a black ellipse in the middle panel of Figure \ref{fig:thinshell_differentangles}). 
We place the two ellipses on the image of NGC 1514 in the lower panel. The simulation produces a structure of two bipolar rings that resembles that of NGC 1514. The simulated ring does not fit exactly the southeast ring. This is because of the departure from mirror symmetry that we discussed in Section \ref{sec:NGC1514} 
Overall, we conclude from this section that non-radiative jet-shell interactions can produce a bipolar pair of rings as observed in NGC 1514, and some other PNE. 

\subsubsection{Effects of Radiative Cooling}
\label{subsubsec:RadiativeCooling}

\begin{figure} 
\centering
\includegraphics[trim=1.5cm 3.0cm 0.0cm 0cm,width=0.5\textwidth]{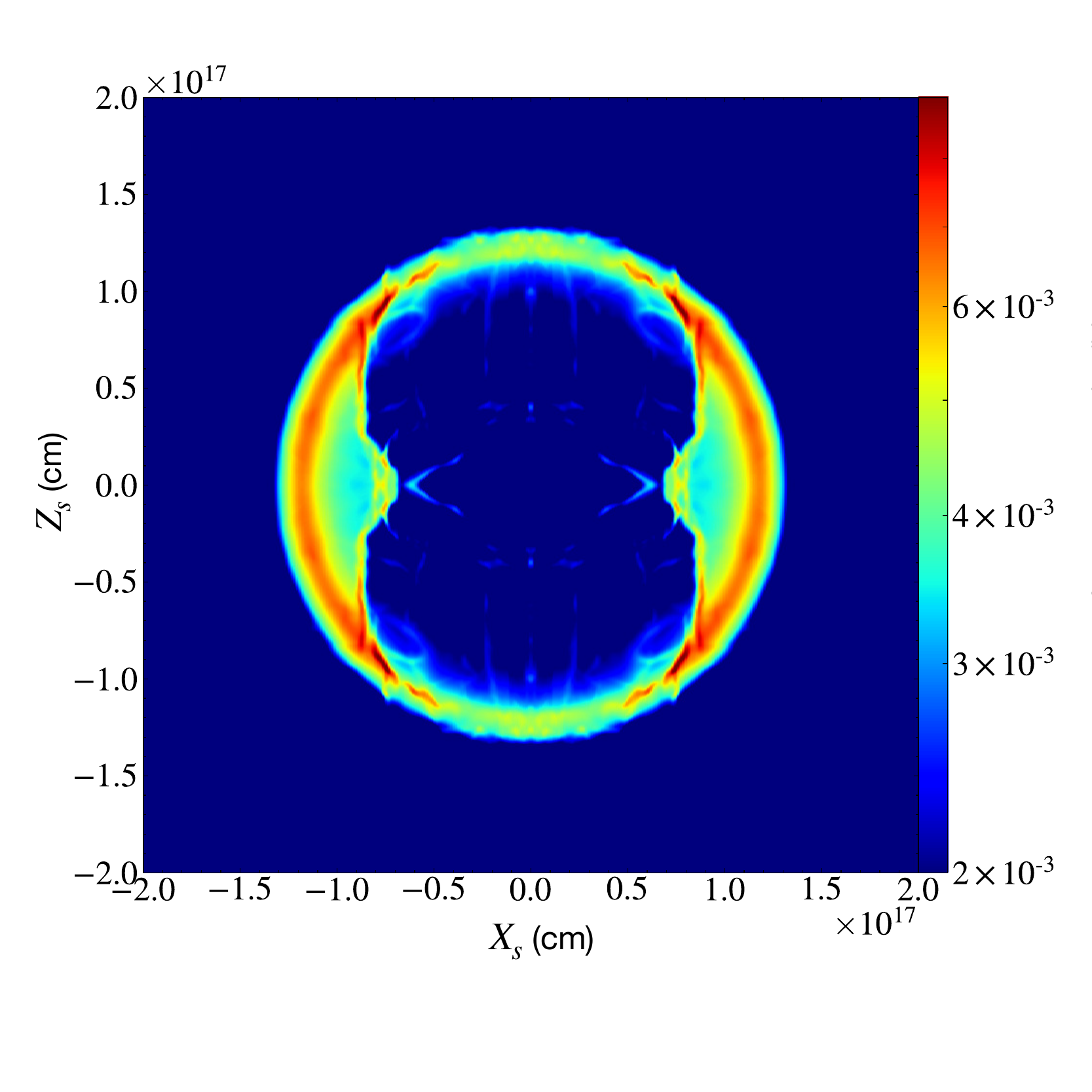}
\caption{A column-density map of a simulation with a radiative cooling of optically thin gas using the cooling function from \cite{Sutherland1993}; all other parameters are as in the non-radiative simulation with a thick dense shell, and at $t=220$ and an inclination of $i=50^\circ$ (compare to the right panel of Figure~\ref{fig:proj_gama167}). The two bipolar rings are very thin and hardly recognizable.}
\label{fig:proj_rad_cool_75}
\end{figure}

Figure~\ref{fig:proj_gama105} of the column density maps for $\gamma=1.05$ presents at late times a bipolar ring structure, but not as prominent as in the non-radiative $\gamma=5/3$ simulations that we present in Figures~\ref{fig:proj_gama167} and \ref{fig:thinshell_differentangles}. 
To better assess the impact of radiative cooling on the ring formation process, we performed an additional simulation including optically thin radiative cooling using the tabulated cooling function of \cite{Sutherland1993}; we kept all other initial conditions and jet parameters as in the non-radiative simulation with the thick dense shell. 
In Figure~\ref{fig:proj_rad_cool_75} we present the column density of this simulation at $t=220 \yr$ and an inclination angle of $i=50^\circ$. 

We learn from Figures~\ref{fig:proj_gama105} and \ref{fig:proj_rad_cool_75} that an efficient radiative cooling suppresses the formation of a prominent bipolar ring structure. There are two rings, but they are thin and hard to observe. It appears that the formation of a central hot bubble, which exerts pressure on the dense shell (compare the density maps in figures \ref{fig:densxz_gama167} and \ref{fig:densxz_gama105}), facilitates ring formation. Other flow parameters may lead to prominent rings even when radiative cooling is efficient, but this requires a large parameter search. 

We expect a non-radiative flow, where the radiative cooling timescale is much longer than the flow timescale, in two limits: either an optically thick flow where the photon diffusion time is much longer than the flow time, or in a low-density, optically thin gas where the radiative cooling time is much longer than the flow time. We discussed these non-radiative limits for jet-shell interaction in \cite{AkashiSoker2013}. Scaling equations from that study, we find the condition for long radiative cooling in an optically thin flow to be that the jet-shell interaction occurs at a large distance of 
\begin{equation}
\begin{split}
     r_{\rm ad} \gtrsim  & 1.6 \times10^4
     \left( \frac{\dot M_{\rm s}}{10^{-4}  M_\odot \yr^{-1}} \right) 
       \left( \frac{v_s}{10 \km \s^{-1}} \right) ^{-1}
       \\       & \times
       \left( \frac{v_j}{800 \km \s^{-1}} \right) ^{-4}
         \left( \frac{\delta}{0.3} \right)^{-1} \AU  ,
     \end{split}
\label{eq:RadLarege}
\end{equation}
where $4 \pi \delta$ is the solid angle covered by the two jets. 
We consider the condition for long radiative cooling in an optically thin flow unlikely in a rapid, violent binary interaction. Instead, we consider the jet-shell interaction to occur under conditions where the photon diffusion time is longer than the expansion time. We take $v_f$ as the flow velocity of the jet-shell interaction structure after the jet-shell interaction, to be somewhere below the jet's initial velocity. The condition for a non-radiative flow is that the interaction takes place very close to the AGB star, at a radius 
\begin{equation}
\begin{split}
     r_{\rm ad} \lesssim  10
     \left( \frac{M_{\rm s}}{0.1 M_\odot} \right)^{1/2} 
       \left( \frac{v_f}{500 \km \s^{-1}} \right) ^{1/2} \AU.
     \end{split}
\label{eq:RadSmall}
\end{equation}
As in \cite{AkashiSoker2013} the opacity is $\kappa=0.34 \cm^2 \g^{-1}$. 

We conclude that the formation of prominent rings requires the jet-shell interaction to take place close to the binary system. In that case, the orbital motion of the binary system has a significant impact that future simulations must take into account. Namely, the interaction might take place on a time scale shorter, or not much longer, than the orbital period. Such an interaction is consistent with a bright transient event, specifically an intermediate luminosity optical transient (ILOT) event. \cite{SokerKashi2012} discussed the formation of some bipolar PNe in ILOT events. The typical interaction timescale will be $\approx 0.01- 0.001$ times the length of our simulations here, namely, about several months to a few years; we can scale our results concerning the shaping of rings to smaller distances and shorter times.  

\section{Discussion and Summary}
\label{sec:Summary}
\subsection{Some earlier studies}
\label{subsec:Earlier}

Many PNe possess equatorial and bipolar ring structures. Understanding the physical processes that shape these structures remains a central question in the late stages of stellar evolution. We take the approach that jets launched by a binary, or tertiary, system shape bipolar rings. There are other models. 
Some early theoretical frameworks demonstrated that magnetized winds and stellar rotation might produce bipolar and equatorial density enhancements. \citet{Garcia1999} used 2D magneto-hydrodynamic (MHD) simulations to show how rotation and magnetic fields in post-AGB winds yield axisymmetric, hourglass-shaped nebulae with collimated polar outflows—structures akin to equatorial rings and bipolar lobes. Extending this to 3D, \citet{Garcia2000} explored point‑symmetric features and misaligned magnetic axes. Subsequently, \citet{Garcia2001} modeled magnetic-pressure oscillations during late AGB phases, showing how periodic field reversals generate concentric shell-like rings. Binary-oriented MHD models, such as \citet{GarciaSegura2018}, further emphasize the interplay of jets, magnetic fields, and companion-driven dynamics.

\citet{HuarteEspinosa2012} modeled the transition from strongly bipolar proto-PNe to more elliptical PNe, highlighting the role of episodic jets and fast winds in shaping ring and lobe structures. \citet{Chen_2016} modeled the L2~Pup system, demonstrating how jet pulses can reproduce a bipolar morphology with internal rings and knots.

The above is a small list of numerical studies out of many more, to demonstrate some different approaches and processes that studies have used, on which we partially based our study. 

Within the framework of jet-shaping, an earlier study by us \citep{Akashietal2015} demonstrated that bipolar jets can compress circumstellar material toward the equator, forming clumpy equatorial rings via Rayleigh–Taylor instabilities (e.g., SN 1987A). In a later paper \citep{Akashi2016}, we explored parameter regimes for the formation of multiple rings. In \cite{Akashi2017Barrel}, we showed that jets interacting with dense spherical shells can produce barrel- and H-shaped nebulae, occasionally with arc-like features reminiscent of rings. In the present study, we also obtained a barrel-shaped structure, as indicated by the density maps at late times (Figures \ref{fig:densxz_gama167} and \ref{fig:densxz_gama105}). The new aim in the present study is to examine a specific PN, NGC 1514, to strengthen the claim for the major role of jets in shaping bipolar rings. 

\subsection{Summary of main results}
\label{subsec:SummaryMainResults}

We focused in this study on the prominent bipolar two-ring structure of NGC 1514, which \cite{Ressleretal2025} analyzed in a recent study. As a side outcome, we found that we can also reproduce the bipolar two-ring structure of Abell~14 (e.g.,  \citealt{Akras2016, Akras2020}.
We performed 3D hydrodynamical simulations of bipolar jets launched into a spherical dense shell. In Section \ref{subsec:Scenario}, we proposed a plausible scenario composed of three phases. In the first phase, the binary interaction enhances the mass loss rate from the AGB progenitor, resulting in the formation of a dense shell. Due to the eccentric orbit and/or the presence of a tertiary object, the shell might have no axial symmetry nor mirror symmetry. In the second phase, the companion accretes mass and launches two opposite jets. We simulated the interaction between the jets launched by the system in the second phase and the shell lost by the system in the first phase. We simulated cases of jets and dense shells with axial and mirror symmetry; however, note that this is a simplification of the actual flow in NGC 1514, as the two rings lack this symmetry (Figures \ref{fig:NGC1514iamges} and \ref{fig:WideRings}). 

We identified a regime of physical parameters that produces two prominent rings in a bipolar structure, similar to the two IR rings of NGC 1514. These parameters are not unique. We can scale our simulations of the jet-shell interaction to occur at much closer or further distances from the binary system. The important parameters are the ratios of jet to shell, mass, velocity, and density. Another very important ratio is the radiative cooling time to the flow time. 

In the case of non-radiative flow interaction, we can find an inclination angle for which the two rings appear similar to the two IR rings of NGC 1514 (Figure~\ref{fig:Comparison}). At a larger angle, namely, almost an edge-on line of sight, the two rings do not cross each other and can resemble the two rings of PN Abell~14 (lower panel of Figure~\ref{fig:thinshell_differentangles}). When radiative cooling is significant, the two rings are much less prominent and might not be observed (Figures~\ref{fig:proj_gama105} and \ref{fig:proj_rad_cool_75}). We expect the shaping to occur at an early phase, as the jet-shell forms close to the binary system and the photon diffusion time is much longer than the flow time (Section \ref{subsubsec:RadiativeCooling}). 

The process of shaping a ring occurs near the jet propagation direction, as the jet interacts with the dense shell material in the polar direction. We could as well form a bipolar double-ring structure when the jets interact with two opposite polar caps instead of a spherical shell. Each polar cap should be significantly wider than the jet, allowing the jet to compress material to the sides. In the case of polar caps, there will be rings but not the barrel-shaped remnant of the dense shell that we find here. 

The short and direct conclusion of this study is that jets interacting with a shell, or polar caps, that the system ejected before launching the jets, can form polar rings. 

We did not simulate the formation of the messy inner structure of NGC 1514. In Section \ref{subsec:Scenario}, we proposed a plausible third phase of interaction, where, in addition to mass accretion from the AGB progenitor, the stellar companion engulfs a sub-stellar companion. The key is that the sub-stellar companion orbital plane around the secondary star is highly inclined to the binary system orbital plane. We proposed that this led to the launching of jets highly inclined to the orbital angular momentum of the binary system. In a previous study \citep{AkashiSoker2017inc}, we demonstrated that inclined jets can lead to the formation of complex planetary nebulae (PNe). 

In future studies, we will simulate all three phases, including the eccentric orbit of the secondary star and the inclined late jets.     

\subsection{Wider implications}
\label{subsec:Implications}
\subsubsection{Circum-jet rings}
\label{subsubsec:CircumJetRings}
We simulated the formation of rings around the axis of the jets. Suppose the jets continue to be active, or there is a later jet-launching episode in the same direction; in that case, we have a structure of circum-jet rings, which are observed in other astrophysical objects, such as active galactic nucleus jets and some core-collapse supernova remnants. While active jets are observed in active galactic nucleus jets, they are not observed in the case of core-collapse supernova remnants and most PNe. \cite{SokerShishkin2025} compared circum-jet rings in one case of active galactic nucleus jets (Cygnus A), two core-collapse supernovae (SNR 0540-69.3 and W49B), and two PNe (MyCn 18 and Hen 2-104). Our results strengthen the claim that jets shape circum-jet rings. This result is particularly important for the core-collapse supernova community, as most members are yet to fully understand the crucial role of jets in the explosion and shaping of core-collapse supernovae and their remnants. 

\subsubsection{Role of jets}
\label{subsubsec:RoleOfJets}

The binary system of NGC 1514 did not experience a common envelope evolution, as its period and eccentricity are $  P\simeq 3300 \days$, and $ e\simeq 0.5$, respectively \citep{Jonesetal2017}. 
We assumed that the companion orbit was sufficiently close to the AGB progenitor of NGC 1514 to accrete mass and launch jets, and it is responsible for the main shaping of NGC 1514. If this is indeed the case, our claim for jet-shaping shows that companions that avoid common envelope evolution can accrete mass and launch energetic jets. 

\cite{Soker2025RAARobust} argued that jets are the most robust observable ingredient of common envelope evolution, and must be included in simulations and calculations to understand the basics of common envelope evolution of many binary systems. We add to that conclusion the possibility that jets also play significant roles in the evolution of binary systems that avoid common envelope evolution and grazing envelope evolution, but experience high mass-transfer rates. Namely, we argue that jets play a significant and even crucial role in binary systems that experience stable mass transfer at high rates.

\section*{Acknowledgements}

We thank an anonymous referee for useful comments. A grant from the Pazy Foundation supported this research. N.S. thanks the Charles Wolfson Academic Chair at the Technion for the support.


\section*{Software}

We use the following software. 
\begin{itemize}
    \item \textbf{FLASH}: A modular, parallel, adaptive-mesh, multi-physics simulation code developed at the University of Chicago, designed to model astrophysical phenomena including compressible, reactive flows \citep{Fryxelletal2000}. 
    \item \textbf{VisIt}: An open-source, interactive, parallel visualization and analysis tool developed by Lawrence Livermore National Laboratory, used here for post-processing and visualization of simulation outputs \citep{Childs2012}.
    \item \textbf{Python}: A high-level programming language employed for supplementary data analysis and plotting, along with the \texttt{NumPy} and \texttt{Matplotlib} libraries \citep{Harris2020,Hunter2007}.
\end{itemize}

\section*{Data and Code Availability}

 The codes used in this article and the data underlying this article will be shared on reasonable request to the corresponding author.


\begin{thebibliography}{}
\makeatletter
\relax
\def\mn@urlcharsother{\let\do\@makeother \do\$\do\&\do\#\do\^\do\_\do\%\do\~}
\def\mn@doi{\begingroup\mn@urlcharsother \@ifnextchar [ {\mn@doi@} {\mn@doi@[]}}
\def\mn@doi@[#1]#2{\def\@tempa{#1}\ifx\@tempa\@empty \href {http://dx.doi.org/#2} {doi:#2}\else \href {http://dx.doi.org/#2} {#1}\fi \endgroup}
\def\mn@eprint#1#2{\mn@eprint@#1:#2::\@nil}
\def\mn@eprint@arXiv#1{\href {http://arxiv.org/abs/#1} {{\tt arXiv:#1}}}
\def\mn@eprint@dblp#1{\href {http://dblp.uni-trier.de/rec/bibtex/#1.xml} {dblp:#1}}
\def\mn@eprint@#1:#2:#3:#4\@nil{\def\@tempa {#1}\def\@tempb {#2}\def\@tempc {#3}\ifx \@tempc \@empty \let \@tempc \@tempb \let \@tempb \@tempa \fi \ifx \@tempb \@empty \def\@tempb {arXiv}\fi \@ifundefined {mn@eprint@\@tempb}{\@tempb:\@tempc}{\expandafter \expandafter \csname mn@eprint@\@tempb\endcsname \expandafter{\@tempc}}}

\bibitem[\protect\citeauthoryear{{Ablimit}}{{Ablimit}}{2024}]{Ablimit2024}
{Ablimit} I.,  2024, \mn@doi [arXiv e-prints] {10.48550/arXiv.2407.03985}, \href {https://ui.adsabs.harvard.edu/abs/2024arXiv240703985A} {p. arXiv:2407.03985}

\bibitem[\protect\citeauthoryear{{Akashi} \& {Soker}}{{Akashi} \& {Soker}}{2013}]{AkashiSoker2013}
{Akashi} M.,  {Soker} N.,  2013, \mn@doi [\mnras] {10.1093/mnras/stt1704}, \href {https://ui.adsabs.harvard.edu/abs/2013MNRAS.436.1961A} {436, 1961}

\bibitem[\protect\citeauthoryear{{Akashi} \& {Soker}}{{Akashi} \& {Soker}}{2016}]{Akashi2016}
{Akashi} M.,  {Soker} N.,  2016, \mn@doi [\mnras] {10.1093/mnras/stw1683}, \href {https://ui.adsabs.harvard.edu/abs/2016MNRAS.462..206A} {462, 206}

\bibitem[\protect\citeauthoryear{{Akashi} \& {Soker}}{{Akashi} \& {Soker}}{2017}]{AkashiSoker2017inc}
{Akashi} M.,  {Soker} N.,  2017, \mn@doi [\mnras] {10.1093/mnras/stx1058}, \href {https://ui.adsabs.harvard.edu/abs/2017MNRAS.469.3296A} {469, 3296}

\bibitem[\protect\citeauthoryear{{Akashi} \& {Soker}}{{Akashi} \& {Soker}}{2018}]{AkashiSoker2018}
{Akashi} M.,  {Soker} N.,  2018, \mn@doi [\mnras] {10.1093/mnras/sty2479}, \href {https://ui.adsabs.harvard.edu/abs/2018MNRAS.481.2754A} {481, 2754}

\bibitem[\protect\citeauthoryear{{Akashi}, {Sabach}, {Yogev}  \& {Soker}}{{Akashi} et~al.}{2015}]{Akashietal2015}
{Akashi} M.,  {Sabach} E.,  {Yogev} O.,   {Soker} N.,  2015, \mn@doi [\mnras] {10.1093/mnras/stv1666}, \href {https://ui.adsabs.harvard.edu/abs/2015MNRAS.453.2115A} {453, 2115}

\bibitem[\protect\citeauthoryear{{Akashi}, {Bear}  \& {Soker}}{{Akashi} et~al.}{2018}]{Akashi2017Barrel}
{Akashi} M.,  {Bear} E.,   {Soker} N.,  2018, \mn@doi [\mnras] {10.1093/mnras/sty029}, \href {https://ui.adsabs.harvard.edu/abs/2018MNRAS.475.4794A} {475, 4794}

\bibitem[\protect\citeauthoryear{{Akras} \& {Gon{\c{c}}alves}}{{Akras} \& {Gon{\c{c}}alves}}{2016}]{AkrasGonclves2016}
{Akras} S.,  {Gon{\c{c}}alves} D.~R.,  2016, \mn@doi [\mnras] {10.1093/mnras/stv2139}, \href {https://ui.adsabs.harvard.edu/abs/2016MNRAS.455..930A} {455, 930}

\bibitem[\protect\citeauthoryear{{Akras}, {Clyne}, {Boumis}, {Monteiro}, {Gon{\c{c}}alves}, {Redman}  \& {Williams}}{{Akras} et~al.}{2016a}]{Akras2016}
{Akras} S.,  {Clyne} N.,  {Boumis} P.,  {Monteiro} H.,  {Gon{\c{c}}alves} D.~R.,  {Redman} M.~P.,   {Williams} S.,  2016a, \mn@doi [\mnras] {10.1093/mnras/stw038}, \href {https://ui.adsabs.harvard.edu/abs/2016MNRAS.457.3409A} {457, 3409}

\bibitem[\protect\citeauthoryear{{Akras}, {Clyne}, {Boumis}, {Monteiro}, {Gon{\c{c}}alves}, {Redman}  \& {Williams}}{{Akras} et~al.}{2016b}]{Akrasetal2016}
{Akras} S.,  {Clyne} N.,  {Boumis} P.,  {Monteiro} H.,  {Gon{\c{c}}alves} D.~R.,  {Redman} M.~P.,   {Williams} S.,  2016b, \mn@doi [\mnras] {10.1093/mnras/stw038}, \href {https://ui.adsabs.harvard.edu/abs/2016MNRAS.457.3409A} {457, 3409}

\bibitem[\protect\citeauthoryear{Akras, Monteiro, Aleman, Farias, May  \& Pereira}{Akras et~al.}{2020}]{Akras2020}
Akras S.,  Monteiro H.,  Aleman I.,  Farias M. A.~F.,  May D.,   Pereira C.~B.,  2020, \mn@doi [Monthly Notices of the Royal Astronomical Society] {10.1093/mnras/staa383}, 493, 2238‚Äì2252

\bibitem[\protect\citeauthoryear{{Aller}, {Montesinos}, {Miranda}, {Solano}  \& {Ulla}}{{Aller} et~al.}{2015}]{Alleretal2015}
{Aller} A.,  {Montesinos} B.,  {Miranda} L.~F.,  {Solano} E.,   {Ulla} A.,  2015, \mn@doi [\mnras] {10.1093/mnras/stv196}, \href {https://ui.adsabs.harvard.edu/abs/2015MNRAS.448.2822A} {448, 2822}

\bibitem[\protect\citeauthoryear{{Avitan} \& {Soker}}{{Avitan} \& {Soker}}{2025}]{AvitanSoker2025}
{Avitan} I.,  {Soker} N.,  2025, \mn@doi [The Open Journal of Astrophysics] {10.33232/001c.131968}, \href {https://ui.adsabs.harvard.edu/abs/2025OJAp....8E..25A} {8, 25}

\bibitem[\protect\citeauthoryear{{Baan}, {Imai}  \& {Orosz}}{{Baan} et~al.}{2021}]{Baanetal2021}
{Baan} W.~A.,  {Imai} H.,   {Orosz} G.,  2021, \mn@doi [Research in Astronomy and Astrophysics] {10.1088/1674-4527/21/11/275}, \href {https://ui.adsabs.harvard.edu/abs/2021RAA....21..275B} {21, 275}

\bibitem[\protect\citeauthoryear{{Balick}}{{Balick}}{1987}]{Balick1987}
{Balick} B.,  1987, \mn@doi [\aj] {10.1086/114504}, \href {https://ui.adsabs.harvard.edu/abs/1987AJ.....94..671B} {94, 671}

\bibitem[\protect\citeauthoryear{{Balick}, {Frank}  \& {Liu}}{{Balick} et~al.}{2020}]{Balicketal2020}
{Balick} B.,  {Frank} A.,   {Liu} B.,  2020, \mn@doi [\apj] {10.3847/1538-4357/ab5651}, \href {https://ui.adsabs.harvard.edu/abs/2020ApJ...889...13B} {889, 13}

\bibitem[\protect\citeauthoryear{{Bandyopadhyay}, {Das}, {Parthasarathy}  \& {Kar}}{{Bandyopadhyay} et~al.}{2023}]{Bandyopadhyayetal2023}
{Bandyopadhyay} R.,  {Das} R.,  {Parthasarathy} M.,   {Kar} S.,  2023, \mn@doi [\mnras] {10.1093/mnras/stad1897}, \href {https://ui.adsabs.harvard.edu/abs/2023MNRAS.524.1547B} {524, 1547}

\bibitem[\protect\citeauthoryear{{Bear} \& {Soker}}{{Bear} \& {Soker}}{2017}]{BearSoker2017Triple}
{Bear} E.,  {Soker} N.,  2017, \mn@doi [\apjl] {10.3847/2041-8213/aa611c}, \href {https://ui.adsabs.harvard.edu/abs/2017ApJ...837L..10B} {837, L10}

\bibitem[\protect\citeauthoryear{{Boffin}, {Miszalski}, {Rauch}, {Jones}, {Corradi}, {Napiwotzki}, {Day-Jones}  \& {K{\"o}ppen}}{{Boffin} et~al.}{2012}]{Boffinetal2012}
{Boffin} H. M.~J.,  {Miszalski} B.,  {Rauch} T.,  {Jones} D.,  {Corradi} R. L.~M.,  {Napiwotzki} R.,  {Day-Jones} A.~C.,   {K{\"o}ppen} J.,  2012, \mn@doi [Science] {10.1126/science.1225386}, \href {https://ui.adsabs.harvard.edu/abs/2012Sci...338..773B} {338, 773}

\bibitem[\protect\citeauthoryear{Chen, Nordhaus, Frank, Blackman  \& Balick}{Chen et~al.}{2016}]{Chen_2016}
Chen Z.,  Nordhaus J.,  Frank A.,  Blackman E.~G.,   Balick B.,  2016, \mn@doi [Monthly Notices of the Royal Astronomical Society] {10.1093/mnras/stw1305}, 460, 4182‚Äì4187

\bibitem[\protect\citeauthoryear{Childs et~al.,}{Childs et~al.}{2012}]{Childs2012}
Childs H.,  et~al., 2012, in Bethel E.~W.,  Childs H.,   Hansen C.,  eds, , High Performance Visualization--Enabling Extreme-Scale Scientific Insight.
CRC Press/Chapman and Hall, pp 357--372

\bibitem[\protect\citeauthoryear{{Chong}, {Kwok}, {Imai}, {Tafoya}  \& {Chibueze}}{{Chong} et~al.}{2012}]{Chongetal2012}
{Chong} S.~N.,  {Kwok} S.,  {Imai} H.,  {Tafoya} D.,   {Chibueze} J.,  2012, \mn@doi [\apj] {10.1088/0004-637X/760/2/115}, \href {https://ui.adsabs.harvard.edu/abs/2012ApJ...760..115C} {760, 115}

\bibitem[\protect\citeauthoryear{{Chu}, {Jacoby}  \& {Arendt}}{{Chu} et~al.}{1987}]{Chuetal1987}
{Chu} Y.-H.,  {Jacoby} G.~H.,   {Arendt} R.,  1987, \mn@doi [\apjs] {10.1086/191207}, \href {https://ui.adsabs.harvard.edu/abs/1987ApJS...64..529C} {64, 529}

\bibitem[\protect\citeauthoryear{{Clairmont}, {Steffen}  \& {Koning}}{{Clairmont} et~al.}{2022}]{Clairmontetal2022}
{Clairmont} R.,  {Steffen} W.,   {Koning} N.,  2022, \mn@doi [\mnras] {10.1093/mnras/stac2375}, \href {https://ui.adsabs.harvard.edu/abs/2022MNRAS.516.2711C} {516, 2711}

\bibitem[\protect\citeauthoryear{{Clark}, {L{\'o}pez}, {Steffen}  \& {Richer}}{{Clark} et~al.}{2013}]{Clarketal2013}
{Clark} D.~M.,  {L{\'o}pez} J.~A.,  {Steffen} W.,   {Richer} M.~G.,  2013, \mn@doi [\aj] {10.1088/0004-6256/145/3/57}, \href {https://ui.adsabs.harvard.edu/abs/2013AJ....145...57C} {145, 57}

\bibitem[\protect\citeauthoryear{{Danehkar}}{{Danehkar}}{2022}]{Danehkar2022}
{Danehkar} A.,  2022, \mn@doi [\apjs] {10.3847/1538-4365/ac5cca}, \href {https://ui.adsabs.harvard.edu/abs/2022ApJS..260...14D} {260, 14}

\bibitem[\protect\citeauthoryear{{Derlopa}, {Akras}, {Amram}, {Boumis}, {Chiotellis}  \& {de Oliveira}}{{Derlopa} et~al.}{2024}]{Derlopaetal2024}
{Derlopa} S.,  {Akras} S.,  {Amram} P.,  {Boumis} P.,  {Chiotellis} A.,   {de Oliveira} C.~M.,  2024, \mn@doi [\mnras] {10.1093/mnras/stae1013}, \href {https://ui.adsabs.harvard.edu/abs/2024MNRAS.530.3327D} {530, 3327}

\bibitem[\protect\citeauthoryear{{Estrella-Trujillo}, {Hern{\'a}ndez-Mart{\'\i}nez}, {Vel{\'a}zquez}, {Esquivel}  \& {Raga}}{{Estrella-Trujillo} et~al.}{2019}]{EstrellaTrujilloetal2019}
{Estrella-Trujillo} D.,  {Hern{\'a}ndez-Mart{\'\i}nez} L.,  {Vel{\'a}zquez} P.~F.,  {Esquivel} A.,   {Raga} A.~C.,  2019, \mn@doi [\apj] {10.3847/1538-4357/ab12e1}, \href {https://ui.adsabs.harvard.edu/abs/2019ApJ...876...29E} {876, 29}

\bibitem[\protect\citeauthoryear{{Fryxell} et~al.,}{{Fryxell} et~al.}{2000}]{Fryxelletal2000}
{Fryxell} B.,  et~al., 2000, \mn@doi [\apjs] {10.1086/317361}, \href {https://ui.adsabs.harvard.edu/abs/2000ApJS..131..273F} {131, 273}

\bibitem[\protect\citeauthoryear{{Garc{\'\i}a-Segura} \& {L{\'o}pez}}{{Garc{\'\i}a-Segura} \& {L{\'o}pez}}{2000}]{Garcia2000}
{Garc{\'\i}a-Segura} G.,  {L{\'o}pez} J.~A.,  2000, \mn@doi [\apj] {10.1086/317186}, \href {https://ui.adsabs.harvard.edu/abs/2000ApJ...544..336G} {544, 336}

\bibitem[\protect\citeauthoryear{{Garc{\'\i}a-Segura}, {Langer}, {R{\'o}{\.z}yczka}  \& {Franco}}{{Garc{\'\i}a-Segura} et~al.}{1999}]{Garcia1999}
{Garc{\'\i}a-Segura} G.,  {Langer} N.,  {R{\'o}{\.z}yczka} M.,   {Franco} J.,  1999, \mn@doi [\apj] {10.1086/307205}, \href {https://ui.adsabs.harvard.edu/abs/1999ApJ...517..767G} {517, 767}

\bibitem[\protect\citeauthoryear{{Garc{\'\i}a-Segura}, {L{\'o}pez}  \& {Franco}}{{Garc{\'\i}a-Segura} et~al.}{2001}]{Garcia2001}
{Garc{\'\i}a-Segura} G.,  {L{\'o}pez} J.~A.,   {Franco} J.,  2001, \mn@doi [\apj] {10.1086/323072}, \href {https://ui.adsabs.harvard.edu/abs/2001ApJ...560..928G} {560, 928}

\bibitem[\protect\citeauthoryear{{Garc{\'\i}a-Segura}, {Ricker}  \& {Taam}}{{Garc{\'\i}a-Segura} et~al.}{2018}]{GarciaSegura2018}
{Garc{\'\i}a-Segura} G.,  {Ricker} P.~M.,   {Taam} R.~E.,  2018, \mn@doi [\apj] {10.3847/1538-4357/aac08c}, \href {https://ui.adsabs.harvard.edu/abs/2018ApJ...860...19G} {860, 19}

\bibitem[\protect\citeauthoryear{{Garc{\'\i}a-Segura}, {Taam}  \& {Ricker}}{{Garc{\'\i}a-Segura} et~al.}{2021}]{GarciaSeguraetal2021}
{Garc{\'\i}a-Segura} G.,  {Taam} R.~E.,   {Ricker} P.~M.,  2021, \mn@doi [\apj] {10.3847/1538-4357/abfc4e}, \href {https://ui.adsabs.harvard.edu/abs/2021ApJ...914..111G} {914, 111}

\bibitem[\protect\citeauthoryear{{Garc{\'\i}a-Segura}, {Taam}  \& {Ricker}}{{Garc{\'\i}a-Segura} et~al.}{2022}]{GarciaSeguraetal2022}
{Garc{\'\i}a-Segura} G.,  {Taam} R.~E.,   {Ricker} P.~M.,  2022, \mn@doi [\mnras] {10.1093/mnras/stac2824}, \href {https://ui.adsabs.harvard.edu/abs/2022MNRAS.517.3822G} {517, 3822}

\bibitem[\protect\citeauthoryear{{Girardi}, {Bressan}, {Bertelli}  \& {Chiosi}}{{Girardi} et~al.}{2000}]{Girardietal2000}
{Girardi} L.,  {Bressan} A.,  {Bertelli} G.,   {Chiosi} C.,  2000, \mn@doi [\aaps] {10.1051/aas:2000126}, \href {https://ui.adsabs.harvard.edu/abs/2000A&AS..141..371G} {141, 371}

\bibitem[\protect\citeauthoryear{{Gold}, {Schmidt}  \& {Ziurys}}{{Gold} et~al.}{2024}]{Goldetal2024}
{Gold} K.~R.,  {Schmidt} D.~R.,   {Ziurys} L.~M.,  2024, \mn@doi [\apj] {10.3847/1538-4357/ad83be}, \href {https://ui.adsabs.harvard.edu/abs/2024ApJ...976..196G} {976, 196}

\bibitem[\protect\citeauthoryear{{G{\'o}mez-Gonz{\'a}lez}, {Toal{\'a}}, {Guerrero}, {Todt}, {Sabin}, {Ramos-Larios}  \& {Mayya}}{{G{\'o}mez-Gonz{\'a}lez} et~al.}{2020}]{GomezGonzalezetal2020}
{G{\'o}mez-Gonz{\'a}lez} V.~M.~A.,  {Toal{\'a}} J.~A.,  {Guerrero} M.~A.,  {Todt} H.,  {Sabin} L.,  {Ramos-Larios} G.,   {Mayya} Y.~D.,  2020, \mn@doi [\mnras] {10.1093/mnras/staa1542}, \href {https://ui.adsabs.harvard.edu/abs/2020MNRAS.496..959G} {496, 959}

\bibitem[\protect\citeauthoryear{{G{\'o}mez-Mu{\~n}oz}, {V{\'a}zquez}, {Sabin}, {Olgu{\'\i}n}, {Guill{\'e}n}, {Zavala}  \& {Michel}}{{G{\'o}mez-Mu{\~n}oz} et~al.}{2023}]{GomezMunozetal2023}
{G{\'o}mez-Mu{\~n}oz} M.~A.,  {V{\'a}zquez} R.,  {Sabin} L.,  {Olgu{\'\i}n} L.,  {Guill{\'e}n} P.~F.,  {Zavala} S.,   {Michel} R.,  2023, \mn@doi [\aap] {10.1051/0004-6361/202346455}, \href {https://ui.adsabs.harvard.edu/abs/2023A&A...676A.101G} {676, A101}

\bibitem[\protect\citeauthoryear{{Guerrero} \& {Manchado}}{{Guerrero} \& {Manchado}}{1998}]{Guerreroetal1998}
{Guerrero} M.~A.,  {Manchado} A.,  1998, \mn@doi [\apj] {10.1086/306407}, \href {https://ui.adsabs.harvard.edu/abs/1998ApJ...508..262G} {508, 262}

\bibitem[\protect\citeauthoryear{{Guerrero}, {Miranda}, {Ramos-Larios}  \& {V{\'a}zquez}}{{Guerrero} et~al.}{2013}]{Guerreroetal2013}
{Guerrero} M.~A.,  {Miranda} L.~F.,  {Ramos-Larios} G.,   {V{\'a}zquez} R.,  2013, \mn@doi [\aap] {10.1051/0004-6361/201220592}, \href {https://ui.adsabs.harvard.edu/abs/2013A&A...551A..53G} {551, A53}

\bibitem[\protect\citeauthoryear{{Guerrero}, {Suzett Rechy-Garc{\'\i}a}  \& {Ortiz}}{{Guerrero} et~al.}{2020}]{Guerreroetal2020}
{Guerrero} M.~A.,  {Suzett Rechy-Garc{\'\i}a} J.,   {Ortiz} R.,  2020, \mn@doi [\apj] {10.3847/1538-4357/ab61fa}, \href {https://ui.adsabs.harvard.edu/abs/2020ApJ...890...50G} {890, 50}

\bibitem[\protect\citeauthoryear{{Guerrero}, {Cazzoli}, {Rechy-Garc{\'\i}a}, {Ramos-Larios}, {Montoro-Molina}, {G{\'o}mez-Gonz{\'a}lez}, {Toal{\'a}}  \& {Fang}}{{Guerrero} et~al.}{2021}]{Guerreoetal2021}
{Guerrero} M.~A.,  {Cazzoli} S.,  {Rechy-Garc{\'\i}a} J.~S.,  {Ramos-Larios} G.,  {Montoro-Molina} B.,  {G{\'o}mez-Gonz{\'a}lez} V.~M.~A.,  {Toal{\'a}} J.~A.,   {Fang} X.,  2021, \mn@doi [\apj] {10.3847/1538-4357/abe2aa}, \href {https://ui.adsabs.harvard.edu/abs/2021ApJ...909...44G} {909, 44}

\bibitem[\protect\citeauthoryear{{Harman}, {Bryce}, {L{\'o}pez}, {Meaburn}  \& {Holloway}}{{Harman} et~al.}{2004}]{Harmanetal2004}
{Harman} D.~J.,  {Bryce} M.,  {L{\'o}pez} J.~A.,  {Meaburn} J.,   {Holloway} A.~J.,  2004, \mn@doi [\mnras] {10.1111/j.1365-2966.2004.07427.x}, \href {https://ui.adsabs.harvard.edu/abs/2004MNRAS.348.1047H} {348, 1047}

\bibitem[\protect\citeauthoryear{Harris et~al.,}{Harris et~al.}{2020}]{Harris2020}
Harris C.~R.,  et~al., 2020, \mn@doi [Nature] {10.1038/s41586-020-2649-2}, 585, 357

\bibitem[\protect\citeauthoryear{{Hillwig}, {Frew}, {Reindl}, {Rotter}, {Webb}  \& {Margheim}}{{Hillwig} et~al.}{2017}]{Hillwigetal2017}
{Hillwig} T.~C.,  {Frew} D.~J.,  {Reindl} N.,  {Rotter} H.,  {Webb} A.,   {Margheim} S.,  2017, \mn@doi [\aj] {10.3847/1538-3881/153/1/24}, \href {https://ui.adsabs.harvard.edu/abs/2017AJ....153...24H} {153, 24}

\bibitem[\protect\citeauthoryear{{Hrivnak}, {Kwok}  \& {Su}}{{Hrivnak} et~al.}{1999}]{Hrivnaketal1999}
{Hrivnak} B.~J.,  {Kwok} S.,   {Su} K. Y.~L.,  1999, \mn@doi [\apj] {10.1086/307822}, \href {https://ui.adsabs.harvard.edu/abs/1999ApJ...524..849H} {524, 849}

\bibitem[\protect\citeauthoryear{{Hsia}, {Chau}, {Zhang}  \& {Kwok}}{{Hsia} et~al.}{2014}]{Hsiaetal2014}
{Hsia} C.-H.,  {Chau} W.,  {Zhang} Y.,   {Kwok} S.,  2014, \mn@doi [\apj] {10.1088/0004-637X/787/1/25}, \href {https://ui.adsabs.harvard.edu/abs/2014ApJ...787...25H} {787, 25}

\bibitem[\protect\citeauthoryear{{Huang}, {Lee}, {Moraghan}  \& {Smith}}{{Huang} et~al.}{2016}]{Huangetal2016}
{Huang} P.-S.,  {Lee} C.-F.,  {Moraghan} A.,   {Smith} M.,  2016, \mn@doi [\apj] {10.3847/0004-637X/820/2/134}, \href {https://ui.adsabs.harvard.edu/abs/2016ApJ...820..134H} {820, 134}

\bibitem[\protect\citeauthoryear{{Huarte-Espinosa}, {Frank}, {Balick}, {Blackman}, {De Marco}, {Kastner}  \& {Sahai}}{{Huarte-Espinosa} et~al.}{2012}]{HuarteEspinosa2012}
{Huarte-Espinosa} M.,  {Frank} A.,  {Balick} B.,  {Blackman} E.~G.,  {De Marco} O.,  {Kastner} J.~H.,   {Sahai} R.,  2012, \mn@doi [\mnras] {10.1111/j.1365-2966.2012.21348.x}, \href {https://ui.adsabs.harvard.edu/abs/2012MNRAS.424.2055H} {424, 2055}

\bibitem[\protect\citeauthoryear{Hunter}{Hunter}{2007}]{Hunter2007}
Hunter J.~D.,  2007, \mn@doi [Computing in Science \& Engineering] {10.1109/MCSE.2007.55}, 9, 90

\bibitem[\protect\citeauthoryear{{Jones}, {Van Winckel}, {Aller}, {Exter}  \& {De Marco}}{{Jones} et~al.}{2017}]{Jonesetal2017}
{Jones} D.,  {Van Winckel} H.,  {Aller} A.,  {Exter} K.,   {De Marco} O.,  2017, \mn@doi [\aap] {10.1051/0004-6361/201730700}, \href {https://ui.adsabs.harvard.edu/abs/2017A&A...600L...9J} {600, L9}

\bibitem[\protect\citeauthoryear{{Jones} et~al.,}{{Jones} et~al.}{2020}]{Jonesetal2020}
{Jones} D.,  et~al., 2020, \mn@doi [\aap] {10.1051/0004-6361/202038778}, \href {https://ui.adsabs.harvard.edu/abs/2020A&A...642A.108J} {642, A108}

\bibitem[\protect\citeauthoryear{{Jones} et~al.,}{{Jones} et~al.}{2022}]{Jonesetal2022}
{Jones} D.,  et~al., 2022, \mn@doi [\mnras] {10.1093/mnras/stab3736}, \href {https://ui.adsabs.harvard.edu/abs/2022MNRAS.510.3102J} {510, 3102}

\bibitem[\protect\citeauthoryear{{Jones}, {Hillwig}  \& {Reindl}}{{Jones} et~al.}{2023}]{Jonesetal2023}
{Jones} D.,  {Hillwig} T.~C.,   {Reindl} N.,  2023, in {Manteiga} M.,  {Bellot} L.,  {Benavidez} P.,  {de Lorenzo-C{\'a}ceres} A.,  {Fuente} M.~A.,  {Mart{\'\i}nez} M.~J.,  {V{\'a}zquez Acosta} M.,   {Dafonte} C.,  eds, Highlights on Spanish Astrophysics XI. p.~216 (\mn@eprint {arXiv} {2304.06355}), \mn@doi{10.48550/arXiv.2304.06355}

\bibitem[\protect\citeauthoryear{{Kohoutek} \& {Hekela}}{{Kohoutek} \& {Hekela}}{1967}]{Kohoutek1967}
{Kohoutek} L.,  {Hekela} J.,  1967, Bulletin of the Astronomical Institutes of Czechoslovakia, \href {https://ui.adsabs.harvard.edu/abs/1967BAICz..18..203K} {18, 203}

\bibitem[\protect\citeauthoryear{{Kwok}}{{Kwok}}{2024}]{Kwok2024}
{Kwok} S.,  2024, \mn@doi [Galaxies] {10.3390/galaxies12040039}, \href {https://ui.adsabs.harvard.edu/abs/2024Galax..12...39K} {12, 39}

\bibitem[\protect\citeauthoryear{{Kwok}, {Chong}, {Hsia}, {Zhang}  \& {Koning}}{{Kwok} et~al.}{2010}]{Kwoketal2010}
{Kwok} S.,  {Chong} S.-N.,  {Hsia} C.-H.,  {Zhang} Y.,   {Koning} N.,  2010, \mn@doi [\apj] {10.1088/0004-637X/708/1/93}, \href {https://ui.adsabs.harvard.edu/abs/2010ApJ...708...93K} {708, 93}

\bibitem[\protect\citeauthoryear{Lee \& Deane}{Lee \& Deane}{2009}]{LeeDeane2009}
Lee D.,  Deane A.~E.,  2009, \mn@doi [Journal of Computational Physics] {10.1016/j.jcp.2008.08.026}, 228, 952

\bibitem[\protect\citeauthoryear{{L{\'o}pez}, {Meaburn}, {Rodr{\'\i}guez}, {V{\'a}zquez}, {Steffen}  \& {Bryce}}{{L{\'o}pez} et~al.}{2000}]{Lopezetal2000}
{L{\'o}pez} J.~A.,  {Meaburn} J.,  {Rodr{\'\i}guez} L.~F.,  {V{\'a}zquez} R.,  {Steffen} W.,   {Bryce} M.,  2000, \mn@doi [\apj] {10.1086/309122}, \href {https://ui.adsabs.harvard.edu/abs/2000ApJ...538..233L} {538, 233}

\bibitem[\protect\citeauthoryear{{Manchado}, {Stanghellini}  \& {Guerrero}}{{Manchado} et~al.}{1996}]{Manchadoetal1996b}
{Manchado} A.,  {Stanghellini} L.,   {Guerrero} M.~A.,  1996, \mn@doi [\apjl] {10.1086/310170}, \href {https://ui.adsabs.harvard.edu/abs/1996ApJ...466L..95M} {466, L95}

\bibitem[\protect\citeauthoryear{{Mellema}}{{Mellema}}{2001}]{Mellema2001}
{Mellema} G.,  2001, \mn@doi [\apss] {10.1023/A:1002737428241}, \href {https://ui.adsabs.harvard.edu/abs/2001Ap&SS.275..147M} {275, 147}

\bibitem[\protect\citeauthoryear{{Mellema}}{{Mellema}}{2003}]{Mellema2003}
{Mellema} G.,  2003, in {Arthur} J.,  {Henney} W.~J.,  eds,  Revista Mexicana de Astronomia y Astrofisica Conference Series Vol. 15, Revista Mexicana de Astronomia y Astrofisica Conference Series. pp 41--43

\bibitem[\protect\citeauthoryear{{Miranda}, {Torrelles}, {Guerrero}, {Aaquist}  \& {Eiroa}}{{Miranda} et~al.}{1998}]{Mirandaetal1998}
{Miranda} L.~F.,  {Torrelles} J.~M.,  {Guerrero} M.~A.,  {Aaquist} O.~B.,   {Eiroa} C.,  1998, \mn@doi [\mnras] {10.1046/j.1365-8711.1998.01611.x}, \href {https://ui.adsabs.harvard.edu/abs/1998MNRAS.298..243M} {298, 243}

\bibitem[\protect\citeauthoryear{{Miranda}, {V{\'a}zquez}, {Olgu{\'\i}n}, {Guill{\'e}n}  \& {Mat{\'\i}as}}{{Miranda} et~al.}{2024}]{Mirandaetal2024}
{Miranda} L.~F.,  {V{\'a}zquez} R.,  {Olgu{\'\i}n} L.,  {Guill{\'e}n} P.~F.,   {Mat{\'\i}as} J.~M.,  2024, \mn@doi [\aap] {10.1051/0004-6361/202348173}, \href {https://ui.adsabs.harvard.edu/abs/2024A&A...687A.123M} {687, A123}

\bibitem[\protect\citeauthoryear{{Moraga Baez}, {Kastner}, {Balick}, {Montez}  \& {Bublitz}}{{Moraga Baez} et~al.}{2023}]{MoragaBaezetal2023}
{Moraga Baez} P.,  {Kastner} J.~H.,  {Balick} B.,  {Montez} R.,   {Bublitz} J.,  2023, \mn@doi [\apj] {10.3847/1538-4357/aca401}, \href {https://ui.adsabs.harvard.edu/abs/2023ApJ...942...15M} {942, 15}

\bibitem[\protect\citeauthoryear{{Morris}}{{Morris}}{1987}]{Morris1987}
{Morris} M.,  1987, \mn@doi [\pasp] {10.1086/132089}, \href {https://ui.adsabs.harvard.edu/abs/1987PASP...99.1115M} {99, 1115}

\bibitem[\protect\citeauthoryear{{Muthu} \& {Anandarao}}{{Muthu} \& {Anandarao}}{2003}]{MuthuAnandarao2003}
{Muthu} C.,  {Anandarao} B.~G.,  2003, \mn@doi [\aj] {10.1086/379552}, \href {https://ui.adsabs.harvard.edu/abs/2003AJ....126.2963M} {126, 2963}

\bibitem[\protect\citeauthoryear{{Rechy-Garc{\'\i}a}, {Pe{\~n}a}  \& {Vel{\'a}zquez}}{{Rechy-Garc{\'\i}a} et~al.}{2019}]{RechyGarciaetal2019}
{Rechy-Garc{\'\i}a} J.~S.,  {Pe{\~n}a} M.,   {Vel{\'a}zquez} P.~F.,  2019, \mn@doi [\mnras] {10.1093/mnras/sty2758}, \href {https://ui.adsabs.harvard.edu/abs/2019MNRAS.482.1163R} {482, 1163}

\bibitem[\protect\citeauthoryear{{Rechy-Garc{\'\i}a}, {Guerrero}, {Duarte Puertas}, {Chu}, {Toal{\'a}}  \& {Miranda}}{{Rechy-Garc{\'\i}a} et~al.}{2020}]{RechyGarciaetal2020}
{Rechy-Garc{\'\i}a} J.~S.,  {Guerrero} M.~A.,  {Duarte Puertas} S.,  {Chu} Y.~H.,  {Toal{\'a}} J.~A.,   {Miranda} L.~F.,  2020, \mn@doi [\mnras] {10.1093/mnras/stz3326}, \href {https://ui.adsabs.harvard.edu/abs/2020MNRAS.492.1957R} {492, 1957}

\bibitem[\protect\citeauthoryear{{Ressler}, {Cohen}, {Wachter}, {Hoard}, {Mainzer}  \& {Wright}}{{Ressler} et~al.}{2010}]{Ressleretal2010}
{Ressler} M.~E.,  {Cohen} M.,  {Wachter} S.,  {Hoard} D.~W.,  {Mainzer} A.~K.,   {Wright} E.~L.,  2010, \mn@doi [\aj] {10.1088/0004-6256/140/6/1882}, \href {https://ui.adsabs.harvard.edu/abs/2010AJ....140.1882R} {140, 1882}

\bibitem[\protect\citeauthoryear{{Ressler}, {Aller}, {Jones}, {Lau}, {Miranda}  \& {Willacy}}{{Ressler} et~al.}{2025}]{Ressleretal2025}
{Ressler} M.~E.,  {Aller} A.,  {Jones} D.,  {Lau} R.~M.,  {Miranda} L.~F.,   {Willacy} K.,  2025, \mn@doi [\aj] {10.3847/1538-3881/adbbcf}, \href {https://ui.adsabs.harvard.edu/abs/2025AJ....169..236R} {169, 236}

\bibitem[\protect\citeauthoryear{{Rubio}, {V{\'a}zquez}, {Ramos-Larios}, {Guerrero}, {Olgu{\'\i}n}, {Guill{\'e}n}  \& {Mata}}{{Rubio} et~al.}{2015}]{Rubioetal2015}
{Rubio} G.,  {V{\'a}zquez} R.,  {Ramos-Larios} G.,  {Guerrero} M.~A.,  {Olgu{\'\i}n} L.,  {Guill{\'e}n} P.~F.,   {Mata} H.,  2015, \mn@doi [\mnras] {10.1093/mnras/stu2201}, \href {https://ui.adsabs.harvard.edu/abs/2015MNRAS.446.1931R} {446, 1931}

\bibitem[\protect\citeauthoryear{{Sabin}, {V{\'a}zquez}, {L{\'o}pez}, {Garc{\'\i}a-D{\'\i}az}  \& {Ramos-Larios}}{{Sabin} et~al.}{2012}]{Sabinetal2012}
{Sabin} L.,  {V{\'a}zquez} R.,  {L{\'o}pez} J.~A.,  {Garc{\'\i}a-D{\'\i}az} M.~T.,   {Ramos-Larios} G.,  2012, \mn@doi [\rmxaa] {10.48550/arXiv.1203.1297}, \href {https://ui.adsabs.harvard.edu/abs/2012RMxAA..48..165S} {48, 165}

\bibitem[\protect\citeauthoryear{{Sahai}}{{Sahai}}{2000}]{Sahai2000}
{Sahai} R.,  2000, \mn@doi [\apjl] {10.1086/312748}, \href {https://ui.adsabs.harvard.edu/abs/2000ApJ...537L..43S} {537, L43}

\bibitem[\protect\citeauthoryear{{Sahai} \& {Trauger}}{{Sahai} \& {Trauger}}{1998}]{SahaiTrauger1998}
{Sahai} R.,  {Trauger} J.~T.,  1998, \mn@doi [\aj] {10.1086/300504}, \href {https://ui.adsabs.harvard.edu/abs/1998AJ....116.1357S} {116, 1357}

\bibitem[\protect\citeauthoryear{{Sahai}, {S{\'a}nchez Contreras}  \& {Morris}}{{Sahai} et~al.}{2005a}]{Sahaietal2005Starfish}
{Sahai} R.,  {S{\'a}nchez Contreras} C.,   {Morris} M.,  2005a, \mn@doi [\apj] {10.1086/426469}, \href {https://ui.adsabs.harvard.edu/abs/2005ApJ...620..948S} {620, 948}

\bibitem[\protect\citeauthoryear{{Sahai}, {Le Mignant}, {S{\'a}nchez Contreras}, {Campbell}  \& {Chaffee}}{{Sahai} et~al.}{2005b}]{Sahaietal2005}
{Sahai} R.,  {Le Mignant} D.,  {S{\'a}nchez Contreras} C.,  {Campbell} R.~D.,   {Chaffee} F.~H.,  2005b, \mn@doi [\apjl] {10.1086/429586}, \href {https://ui.adsabs.harvard.edu/abs/2005ApJ...622L..53S} {622, L53}

\bibitem[\protect\citeauthoryear{{Sahai}, {Morris}  \& {Villar}}{{Sahai} et~al.}{2011}]{Sahaietal20112011}
{Sahai} R.,  {Morris} M.~R.,   {Villar} G.~G.,  2011, \mn@doi [\aj] {10.1088/0004-6256/141/4/134}, \href {https://ui.adsabs.harvard.edu/abs/2011AJ....141..134S} {141, 134}

\bibitem[\protect\citeauthoryear{{Sahai}, {Vlemmings}  \& {Nyman}}{{Sahai} et~al.}{2017}]{Sahaietal2017}
{Sahai} R.,  {Vlemmings} W.~H.~T.,   {Nyman} L.~{\r{A}}.,  2017, \mn@doi [\apj] {10.3847/1538-4357/aa6d86}, \href {https://ui.adsabs.harvard.edu/abs/2017ApJ...841..110S} {841, 110}

\bibitem[\protect\citeauthoryear{{Sahai} et~al.,}{{Sahai} et~al.}{2024}]{Sahaietal2024}
{Sahai} R.,  et~al., 2024, arXiv e-prints, \href {https://ui.adsabs.harvard.edu/abs/2024arXiv240906038S} {p. arXiv:2409.06038}

\bibitem[\protect\citeauthoryear{{Soker}}{{Soker}}{1990}]{Soker1990AJ}
{Soker} N.,  1990, \mn@doi [\aj] {10.1086/115465}, \href {https://ui.adsabs.harvard.edu/abs/1990AJ.....99.1869S} {99, 1869}

\bibitem[\protect\citeauthoryear{{Soker}}{{Soker}}{2004}]{Soker2004triple}
{Soker} N.,  2004, \mn@doi [\mnras] {10.1111/j.1365-2966.2004.07731.x}, \href {https://ui.adsabs.harvard.edu/abs/2004MNRAS.350.1366S} {350, 1366}

\bibitem[\protect\citeauthoryear{{Soker}}{{Soker}}{2023}]{Soker2025Bright}
{Soker} N.,  2023, \mn@doi [arXiv e-prints] {10.48550/arXiv.2310.15785}, \href {https://ui.adsabs.harvard.edu/abs/2023arXiv231015785S} {p. arXiv:2310.15785}

\bibitem[\protect\citeauthoryear{{Soker}}{{Soker}}{2025}]{Soker2025RAARobust}
{Soker} N.,  2025, \mn@doi [Research in Astronomy and Astrophysics] {10.1088/1674-4527/adb15b}, \href {https://ui.adsabs.harvard.edu/abs/2025RAA....25b5023S} {25, 025023}

\bibitem[\protect\citeauthoryear{{Soker} \& {Kashi}}{{Soker} \& {Kashi}}{2012}]{SokerKashi2012}
{Soker} N.,  {Kashi} A.,  2012, \mn@doi [\apj] {10.1088/0004-637X/746/1/100}, \href {https://ui.adsabs.harvard.edu/abs/2012ApJ...746..100S} {746, 100}

\bibitem[\protect\citeauthoryear{{Soker} \& {Shishkin}}{{Soker} \& {Shishkin}}{2025}]{SokerShishkin2025}
{Soker} N.,  {Shishkin} D.,  2025, \mn@doi [\pasa] {10.1017/pasa.2025.39}, \href {https://ui.adsabs.harvard.edu/abs/2025PASA...42...48S} {42, e048}

\bibitem[\protect\citeauthoryear{{Sowicka}, {Jones}, {Corradi}, {Wesson}, {Garc{\'\i}a-Rojas}, {Santander-Garc{\'\i}a}, {Boffin}  \& {Rodr{\'\i}guez-Gil}}{{Sowicka} et~al.}{2017}]{Sowickaetal2017}
{Sowicka} P.,  {Jones} D.,  {Corradi} R. L.~M.,  {Wesson} R.,  {Garc{\'\i}a-Rojas} J.,  {Santander-Garc{\'\i}a} M.,  {Boffin} H. M.~J.,   {Rodr{\'\i}guez-Gil} P.,  2017, \mn@doi [\mnras] {10.1093/mnras/stx1697}, \href {https://ui.adsabs.harvard.edu/abs/2017MNRAS.471.3529S} {471, 3529}

\bibitem[\protect\citeauthoryear{{Steffen}, {Koning}, {Esquivel}, {Garc{\'\i}a-Segura}, {Garc{\'\i}a-D{\'\i}az}, {L{\'o}pez}  \& {Magnor}}{{Steffen} et~al.}{2013}]{Steffenetal2013}
{Steffen} W.,  {Koning} N.,  {Esquivel} A.,  {Garc{\'\i}a-Segura} G.,  {Garc{\'\i}a-D{\'\i}az} M.~T.,  {L{\'o}pez} J.~A.,   {Magnor} M.,  2013, \mn@doi [\mnras] {10.1093/mnras/stt1583}, \href {https://ui.adsabs.harvard.edu/abs/2013MNRAS.436..470S} {436, 470}

\bibitem[\protect\citeauthoryear{{Sutherland} \& {Dopita}}{{Sutherland} \& {Dopita}}{1993a}]{SutherlandDopita1993}
{Sutherland} R.~S.,  {Dopita} M.~A.,  1993a, \mn@doi [\apjs] {10.1086/191823}, \href {https://ui.adsabs.harvard.edu/abs/1993ApJS...88..253S} {88, 253}

\bibitem[\protect\citeauthoryear{{Sutherland} \& {Dopita}}{{Sutherland} \& {Dopita}}{1993b}]{Sutherland1993}
{Sutherland} R.~S.,  {Dopita} M.~A.,  1993b, \mn@doi [\apjs] {10.1086/191823}, \href {https://ui.adsabs.harvard.edu/abs/1993ApJS...88..253S} {88, 253}

\bibitem[\protect\citeauthoryear{{Tafoya}, {Orosz}, {Vlemmings}, {Sahai}  \& {P{\'e}rez-S{\'a}nchez}}{{Tafoya} et~al.}{2019}]{Tafoyaetal2019}
{Tafoya} D.,  {Orosz} G.,  {Vlemmings} W.~H.~T.,  {Sahai} R.,   {P{\'e}rez-S{\'a}nchez} A.~F.,  2019, \mn@doi [\aap] {10.1051/0004-6361/201834632}, \href {https://ui.adsabs.harvard.edu/abs/2019A&A...629A...8T} {629, A8}

\bibitem[\protect\citeauthoryear{Toro, Spruce  \& Speares}{Toro et~al.}{1994}]{Toro1994}
Toro E.~F.,  Spruce M.,   Speares W.,  1994, \mn@doi [Shock Waves] {10.1007/BF01414629}, 4, 25

\bibitem[\protect\citeauthoryear{{Trammell} \& {Goodrich}}{{Trammell} \& {Goodrich}}{2002}]{Trammelletal2002}
{Trammell} S.~R.,  {Goodrich} R.~W.,  2002, \mn@doi [\apj] {10.1086/342943}, \href {https://ui.adsabs.harvard.edu/abs/2002ApJ...579..688T} {579, 688}

\bibitem[\protect\citeauthoryear{{Ubertosi} et~al.,}{{Ubertosi} et~al.}{2025}]{Ubertosietal2025}
{Ubertosi} F.,  et~al., 2025, \mn@doi [\aap] {10.1051/0004-6361/202452430}, \href {https://ui.adsabs.harvard.edu/abs/2025A&A...693A.171U} {693, A171}

\bibitem[\protect\citeauthoryear{{Vel{\'a}zquez}, {Raga}, {Riera}, {Steffen}, {Esquivel}, {Cant{\'o}}  \& {Haro-Corzo}}{{Vel{\'a}zquez} et~al.}{2012}]{Velazquez2012}
{Vel{\'a}zquez} P.~F.,  {Raga} A.~C.,  {Riera} A.,  {Steffen} W.,  {Esquivel} A.,  {Cant{\'o}} J.,   {Haro-Corzo} S.,  2012, \mn@doi [\mnras] {10.1111/j.1365-2966.2011.19991.x}, \href {https://ui.adsabs.harvard.edu/abs/2012MNRAS.419.3529V} {419, 3529}

\bibitem[\protect\citeauthoryear{{Wen}, {Hsia}, {Kang}, {Chen}  \& {Luo}}{{Wen} et~al.}{2023}]{Wenetal2023}
{Wen} S.-B.,  {Hsia} C.-H.,  {Kang} X.-X.,  {Chen} R.,   {Luo} T.,  2023, \mn@doi [Research in Astronomy and Astrophysics] {10.1088/1674-4527/acbe95}, \href {https://ui.adsabs.harvard.edu/abs/2023RAA....23c5018W} {23, 035018}

\bibitem[\protect\citeauthoryear{{Wen}, {Wang}, {Hsia}, {Yeh}, {Liu}, {Liu}  \& {Kang}}{{Wen} et~al.}{2024}]{Wenetal2024}
{Wen} S.,  {Wang} Y.-Z.,  {Hsia} C.-H.,  {Yeh} S.,  {Liu} J.-Z.,  {Liu} H.-X.,   {Kang} X.-X.,  2024, \mn@doi [\aap] {10.1051/0004-6361/202449751}, \href {https://ui.adsabs.harvard.edu/abs/2024A&A...687A..99W} {687, A99}

\makeatother
\end{thebibliography}

\end{document}